\documentclass{ws-procs9x6}
\renewcommand\draftnote{\hbox to \trimwidth{\empty}}%
\begin{document}
\begin{flushright}
DESY 07-212\\
December 2007
\end{flushright}

\title{Theoretical Interest in $B$-Meson Physics at the B Factories, Tevatron
and the LHC\footnote{Invited Talk, International Symposium on Contemporary
 Physics,
National Centre for Physics, Islamabad,
Pakistan (March 26-30, 2007); to be published in the symposium proceedings.}}

\author{Ahmed Ali}

\address{Deutsches Elektronen-Synchrotron DESY, Hamburg,
Germany\\E-mail: ahmed.ali@desy.de}

\begin{abstract}
We review the salient features of $B$-meson physics, with particular emphasis
on the measurements carried out at the $B$-factories and Tevatron,
 theoretical progress in understanding these measurements in the context of
 the standard model, and  anticipation at the LHC. Topics discussed
 specifically are the
current status of the Cabibbo-Kobayashi-Maskawa matrix, the CP-violating
phases, rare radiative and semileptonic decays, and some selected
non-leptonic two-body decays of the $B$ mesons. 
\end{abstract}

\bodymatter

\section{Introduction}

The two $B$-meson factories operating at the
KEK and SLAC $e^+e^-$ storage rings have outperformed their projected luminosities and have
produced a wealth of data on the process $e^+ e^- \to \Upsilon(4S) \to B
\bar{B}$, with subsequent weak decays of the $B$ and $\bar{B}$ mesons. The
exclusivity of the final state, the asymmetric beam energies
 and the large statistics collected at the $\Upsilon(4S)$ ($O(10^9) B\bar{B}$ events)
have led to a number of impressive and
quantitative results, which include, among other measurements, precise
determination  of the weak mixing (Cabibbo-Kobayashi-Maskawa
 CKM~\cite{Cabibbo:yz,Kobayashi:fv}) matrix
elements
$\vert V_{cb}\vert$ and $\vert V_{ub} \vert$ ~\cite{Barberio-LP07}, a large number of CP-violating
asymmetries
in exclusive decays of the $B$ mesons~\cite{Brown-LP07}, and rare radiative and
semileptonic decays (implying flavor
changing neutral current FCNC transitions)
 involving the $B^0$ and $B^+$ mesons and their charge
 conjugates~\cite{Nakao-LP07}. Recently, Belle
 collaboration have published a number of interesting results on the decays of
 the $B_s$-meson highlighted by the first measurement of the radiative
 penguin decay $B_s \to \phi \gamma$~\cite{Wicht:2007rt}. This is the
 SU(3)-counterpart to the
 much celebrated decay $B \to K^* \gamma$, first measured  by CLEO
and further consolidated by the BABAR and BELLE collaborations~\cite{hfag07}. 

 The two B-meson factories are not the
only players currently active in the study of $B$-meson physics. 
In fact, $b$-physics is the main beneficiary of the upgrade program
at the high energy proton-antiproton collider Tevatron at Fermilab.
Since this upgrade, physics news from Fermilab are dominated by 
the achievements in the $b$-quark sector. We have seen very
impressive and seminal results in this field by the two
Fermilab experiments CDF and D0, including the landmark measurement of the
$B_s^0$ - $\overline{B_s^0}$-mixing induced mass difference $\Delta M_{B_s}$ 
~\cite{Abazov:2006dm,Abulencia:2006mq} and
the first result on direct CP violation in the $B_s$-meson sector ${\cal
  A}_{\rm CP}(B_s^0 \to K^+\pi^-)$
~\cite{Morello:2006pv,Abulencia:2006psa}. These experiments
 have established 
(if any proof was needed) that
cutting edge  flavor physics is done also at hadron machines. In all
likelihood, this success
story of the hadron machines will be set forth at the Large Hadron Collider (LHC), commissioned to
operate in 2008 at CERN. In particular, the  
experiment LHCb~\cite{LHCb-07}, dedicated to precision measurements of the
physics of the entire family
of $B$-mesons ($B^0$, $B^+$, $B_s^0$, $B_c^+$, their charged conjugated states,
and excited states) and $\Lambda_b$ baryons (including the entire baryon
spectrum containing at least one $b$-quark), will greatly broaden our knowledge
of both the spectroscopic and dynamical aspects of $b$-physics.
The other two LHC experiments ATLAS and CMS will also contribute to
$b$ physics. 

In this talk, I will briefly review the highlights of the measurements already
accomplished,
making contacts with theoretical expectations in the standard model (SM). 
The topics discussed are: (i) an update of the CKM matrix, with emphasis on
the matrix elements in the third row and the third column of $V_{\rm CKM}$ from direct decays
(which determine $\vert V_{cb}\vert, \vert V_{ub} \vert$ and  $\vert V_{tb}\vert$),
and from induced transitions involving the mass differences $\Delta M_{B_d}$,
 $\Delta M_{B_s}$ and
the electromagnetic
penguin decays $b \to (s,d)\gamma$ (which determine the matrix elements
$\vert V_{td}\vert$ and $\vert V_{ts}\vert$)  (ii) measurements of
the CP-violating phases $\alpha$, $\beta$ and $\gamma$ from exclusive $B$-decays,
and (iii) radiative,
semileptonic and leptonic rare $B$-decays, a field which has received a lot of
theoretical attention both in the context of the SM and in extensions of it. Much more detailed discussions of
these topics can be found in the original literature cited below, in the
reviews, such as the Review of Particle Properties by  
the Particle Data Group~\cite{Yao:2006px}, and in the proceedings of the 
topical workshops on flavor physics~\cite{Ricciardi:2006yx}.

\section{Status of the CKM Matrix}
The CKM matrix is written below in the
Wolfenstein parameterization~\cite{Wolfenstein:1983yz} in terms of the 
four parameters $A, ~\lambda,~~\rho, ~~\eta$:
\begin{small}
\begin{eqnarray}
V_{\rm CKM} &\equiv& 
\left(
\begin{matrix}
 1-{1\over 2}\lambda^2 & \lambda
 & \hspace*{-0.7cm}A\lambda^3 \left( \rho - i\eta \right) \cr
 -\lambda ( 1 + i A^2 \lambda^4 \eta )
& 1-{1\over 2}\lambda^2 &  \hspace*{-0.7cm}A\lambda^2 \cr
 A\lambda^3\left(1 - \rho - i \eta\right) & -A\lambda^2\left(1+i \lambda^2 \eta\right) 
& \hspace*{-0.7cm}1
\end{matrix}
\right)~. 
\label{CKM-W}
\end{eqnarray}
\end{small}
 Anticipating precision data, a perturbatively improved version~\cite{Buras:1994ec} of
the Wolfenstein parameterization will be used below with
 $\bar \rho =\rho(1-\lambda^2/2),~~\bar \eta= \eta(1-\lambda^2/2)$.

%
\begin{figure}[htbp]
\centerline{\psfig{width=0.60\textwidth,file=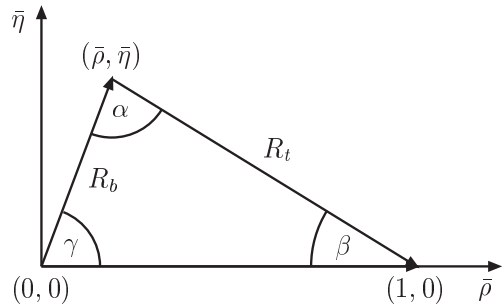}}
\caption{The unitarity triangle with unit base 
in the $\bar{\rho}$ - $\bar{\eta}$ plane.}
\label{triangle}
\end{figure}

 Unitarity of the CKM matrix
implies six  relations, of which the one resulting from the
equation $V_{ud}V_{ub}^* + V_{cd}V_{cb}^* + V_{td} V_{tb}^*=0$ is the principal focus of the
current experiments in $B$-decays.
This is a triangle relation in the complex plane 
(i.e.\ $\bar{\rho}$--$\bar{\eta}$ space), and  the three angles of this 
triangle are called $\alpha$, $\beta$ and $\gamma$, with 
the BELLE convention being $\phi_1=\beta$,  $\phi_2=\alpha$ and 
$\phi_3=\gamma$ (see Fig.\ref{triangle}). The unitarity relation in discussion can also be written as
\begin{equation}
R_b {\rm e}^{i\gamma} + R_t {\rm e}^{-i \beta} =1\,,
\label{eq:trianglerel} 
\end{equation}
where
$R_b = \left (1-
\frac{\lambda^2}{2}\right )\, \frac{1}{\lambda} \,\left \vert 
\frac{V_{ub}}{V_{cb}} \right \vert 
=\sqrt{\bar{\rho}^2 + \bar{\eta}^2}$ and
$R_t =\frac{1}{\lambda} \, \left \vert 
\frac{V_{td}}{V_{cb}} \right \vert =\sqrt{(1-\bar{\rho})^2 + \bar{\eta}^2}$.
Thus, precise determination of $\vert V_{cb}\vert$, $\vert V_{ub} \vert$ and $\vert V_{td} \vert$
and the three CP-violating phases $\alpha$, $\beta$, $\gamma$  is
crucial in testing the CKM paradigm.

In this section, we mainly review 
 the current status of the CKM matrix elements in the third
row and the third column, namely $V_{ub}, V_{cb}, V_{td}, V_{ts}$ and $V_{tb}$. 
The other four CKM matrix elements, $V_{ud},~V_{us},~V_{cd}$ and $ V_{cs}$
 are discussed 
 in~\cite{Yao:2006px,Barberio-LP07}, from where  
 we quote the current best measurements.
\begin{itemize}
\item $V_{ud}$ is obtained from the nuclear $\beta$-decay of the (super-allowed)
  $O^+ \to O^+$ transitions~\cite{Hardy:2004dm,Savard:2005cv}, yielding:
  $V_{ud} =0.97377 (27)$. 
\item $V_{us}$ is obtained from the $K_{\ell 3}$ decays and  chiral
 perturbation theory, yielding $ V_{us} =0.22535 \pm 0.00116$. In getting this, 
 the Leutwyler-Roos value for the form factor $f_+(0)= 0.961 \pm
 0.008$~\cite{Leutwyler:1984je} has been used.
 This is to be contrasted with the value, $V_{us}=\lambda=0.2265(7)$, obtained
 from the unitarity condition on
the first row of the CKM matrix and the input value for $V_{ud}$ given above.
They are in good agreement with each other, providing a precision test of the
 unitarity of the first row of $V_{\rm CKM}$.

\item $V_{cs}$ is obtained from the decay $D \to K e \nu_e$ and the Lattice-QCD result for
  the $D \to K$ form factor~\cite{Kronfeld:2006sk}, yielding
 $\vert V_{cs}\vert =0.996 \pm 0.008 \pm 0.015 \pm 0.104$, where the first two
errors are experimental and the third (dominant) error is from theory. 
\item Despite impressive progress in the measurement of the Cabibbo-suppressed
  $D$-meson decays and Lattice-QCD, the determination of
$V_{cd}$ is still dominated by the neutrino-nucleon production of the charm
  quark, yielding: $\vert V_{cd}\vert =0.230 \pm
  0.011$~\cite{Yao:2006px}. Within errors, $\vert V_{cd}\vert = \vert V_{us}
  \vert$, as anticipated from eq.~(\ref{CKM-W}).
\end{itemize}

\subsection{Current determinations of  $\vert V_{cb}\vert$ and  $\vert
  V_{ub}\vert$}
 Determinations of $\vert V_{cb}\vert $ are
based on the semileptonic decay  $b \to c \ell \nu_\ell$. This
transition can be measured either inclusively through the process $B \to X_c \ell \nu_\ell$, where
$X_c$ is a hadronic state with a net $c$-quantum number, or exclusively, such as the decays
$B \to (D,D^*) \ell \nu_\ell$. In either case, intimate knowledge of QCD is required to go from
the partonic process to the hadronic states. The fact that $m_b \gg \Lambda_{\rm QCD}$ has led to
novel applications of QCD in which heavy quark expansion (HQE) plays a central
 role~\cite{Manohar:dt} and the underlying theory
is termed as HQET.

\subsubsection{$\vert V_{cb} \vert$ from the decays $B~\to~X_c~\ell~\nu_\ell$}
\label{subsec:vcbincl}
 Concentrating first on the inclusive decays,
the semileptonic decay rate can be calculated as a power series
\begin{equation}
\Gamma = \Gamma_0 + \frac{1}{m_b} \Gamma_1 + \frac{1}{m_b^2} \Gamma_2 + \frac{1}{m_b^3} \Gamma_3
+...,  
\end{equation}
where each $\Gamma_i$ is a perturbation series in $\alpha_s(m_b)$, the QCD coupling constant
at the scale $m_b$. Here $\Gamma_0 $
is the decay width of a free $b$-quark, which gives the parton-model result. The coefficient of
the leading power correction $\Gamma_1 $ is absent~\cite{CGG}, and the
effect of the $1/m_b^2$ correction is collected in $\Gamma_2$, which can be expressed in terms of two
non-perturbative parameters called $\lambda_1$ - kinetic energy of the $b$-quark - and
$\lambda_2$ - its chromomagnetic moment. These quantities, also called
$\mu_\pi^2$ and $\mu_G^2$, respectively, in the literature, are defined in terms of the following
matrix elements~\cite{incl,MaWi,Blok:1993va,Mannel:1993su}:
\begin{eqnarray}
2 M_B \lambda_1 &\equiv& \langle B (v) 
                | \bar{Q}_v  (iD)^2  Q_v|  B (v) \rangle ,\\
6 M_B \lambda_2  &\equiv&
\langle B (v)   | \bar{Q}_v \sigma_{\mu \nu} [iD^\mu , iD^\nu]
Q_v | B (v) \rangle~,
 \nonumber
\end{eqnarray}
where $D_\mu$ is the covariant derivative and heavy quark fields are characterized by
the 4-velocity, $v$.
At ${\cal O} (\Lambda_{\rm QCD}^3/m_b^3)$,  six new matrix elements enter in $\Gamma_3 $,
usually denoted by $\rho_{1,2}$ and ${\cal T}_{1,2,3,4}$.

Data have been analyzed in the theoretical accuracy in which corrections up to 
${\cal O}(\alpha_s^2 \beta_0)$, ${\cal O}(\alpha_s \Lambda_{\rm QCD}/m_b)$ and 
${\cal O}(\Lambda_{\rm QCD}^3/m_b^3)$  are taken into
 account~\cite{Bauer:2004ve,Gambino:2004qm}, with $\beta_0$ being the lowest
 order coefficient in the expansion of the QCD
 $\beta$-function.
  In addition to the other parameters, a quark mass scheme has to be specified.
 Bauer et al.~\cite{Bauer:2004ve} have carried out a comprehensive
 study of the scheme dependence using
five quark mass schemes: $1S$, PS, $\overline{\rm MS}$, kinematic, and the pole mass.

To extract the value of $\vert V_{cb}\vert$ and other fit parameters,  three different
distributions, namely the charged lepton energy spectrum and the hadronic invariant mass spectrum in
$B \to X_c \ell \bar {\nu}_\ell$, and the photon energy spectrum in $B \to X_s \gamma$ have
been studied.
Theoretical analyses are carried out in terms of the moments.
 Defining the integral
$R_n(E_{\rm cut}, \mu) \equiv \int_{E_{\rm cut}} dE_\ell (E_\ell -\mu)^n d\Gamma/dE_\ell$,
where $E_{\rm cut}$ is a lower cut on the charged lepton energy, moments of
the lepton energy spectrum are given by $\langle E_\ell^n \rangle = R_n(E_{\rm cut}, 0)/R_0(E_{\rm cut},0)$.
For the $B \to X_c \ell \bar{\nu}_\ell$ hadronic invariant mass spectrum, 
the moments are defined likewise with the cutoff $E_{\rm cut}$.
Analyses of the  data along these lines have been presented  by 
a large number of experiments. For a
summary and references to the original literature, see HFAG~\cite{hfag07}.  
 
The  global fit of the world data
 in the so-called $1S$-scheme for the $b$-quark mass 
undertaken  by Bauer et al.~\cite{Bauer:2004ve}
leads to the following fit values for $\vert V_{cb} \vert$ and $m_b^{1S}$:
\begin{eqnarray}
& &  |V_{cb}| = \left( 41.78 \pm 0.30 \pm 0.08_{\tau_B}\right) \times 10^{-3}, 
\nonumber \\
& &  m_b^{1S} = (4.701 \pm 0.030)\,\mbox{GeV}.
\label{eq:vcb-bauer}
\end{eqnarray}
The  BABAR collaboration have studied the dependence of the lepton and hadron moments
on the cutoff $E_{\rm cut}$ and compared their measurements with the theoretical calculation
by Gambino and Uraltsev~\cite{Gambino:2004qm} using the so-called kinematic scheme for the $b$-quark mass
 $m_b^{\rm kin} (\mu)$, renormalized at the scale $\mu=1$ GeV.
 Agreement between experiment and theory
 allows to determine the fit parameters in this scheme with the
 results~\cite{hfag07}:  
\begin{eqnarray}
\nonumber
\vspace*{-24mm}
\hspace*{-15mm} & &  |V_{cb}| = 
(41.91 \pm 0.19_{exp} \pm 0.28_{HQE} \pm 0.59_{\Gamma_{\rm sl}})\, \times 10^{-3},  
\nonumber \\
& &  m_b~(1 {\rm GeV}) = 
 \hspace*{1mm} 
(4.613 \pm 0.022_{exp} \pm 0.027_{HQE})~{\rm GeV}, 
\nonumber \\
& &  m_c~(1 {\rm GeV}) = 
\hspace*{1mm} 
(1.187 \pm 0.033_{exp} \pm 0.040_{HQE})~{\rm GeV}~.
\label{eq:vcb-babar}
\end{eqnarray}
\vspace*{-1mm}
The two analyses (\ref{eq:vcb-bauer}) and (\ref{eq:vcb-babar}) are in excellent agreement
 with each other. The achieved accuracy
$\delta \vert V_{cb}\vert/\vert V_{cb} \vert \simeq 2\%$ is impressive, and
 the precision on $m_b$ is also remarkable,
$\delta m_b/m_b =O(10^{-3})$, with a similar precision obtained on the mass difference $m_b-m_c$.
\subsection{$\vert V_{cb} \vert$ from $B \to(D,D^*) \ell \nu_\ell$ decays}
\label{subsec:vcbexcl}
The classic application of HQET in heavy $\to$ heavy decays is in the decay 
$B \to D^* \ell \nu_\ell$. The differential distribution in the variable $\omega (=v_B.v_{D^*})$,
where $v_B (v_{D^*})$ is the four-velocity of the $B(D^*)$-meson, is given by
\begin{eqnarray}
\frac{d\Gamma}{d\omega} &=& \frac{G_F^2}{4 \pi^3}  \,  \vert V_{cb} \vert^2 \, m_{D^*}^3 \, 
(m_B - m_{D^*})^2 
 \left (\omega^2 -1 \right )^{1/2} \,{\cal G}(\omega) \, \vert {\cal F} (\omega) \vert^2~,
\nonumber
\end{eqnarray}
where ${\cal G}(\omega)$ is a phase space factor with 
 ${\cal G}(1)=1$, and 
 ${\cal F}(\omega)$ is the Isgur--Wise (IW) function~\cite{Isgur:vq}
 with the normalization at the symmetry point
${\cal F}(1)=1$.  Leading $\Lambda_{\rm QCD}/m_b$ corrections in ${\cal F}(1)$  
 are absent  due to Luke's theorem~\cite{Luke:1990eg}. Theoretical issues are the precise determination of the 
second order power correction to ${\cal F}(\omega=1)$, the
slope  $\rho^2$ and the curvature $c$ of the IW-function:
\begin{equation}
{\cal F}(\omega) ={\cal F}(1)\, \left [1 - \rho^2\,(\omega -1) + 
c\,(\omega -1)^2 + ...\right ].
\nonumber
\end{equation}
Bounds on $\rho^2$ have been obtained by Bjorken~\cite{Bjorken:1990hs} and Uraltsev~\cite{Uraltsev:2000ce},
which can be combined to yield $\rho^2 > 3/4$. Likewise,
 bounds on the second (and higher) derivatives of the IW-function have been worked out
 by the Orsay group~\cite{Jugeau:2004qd}, yielding $c > 15/32$.
 Apparently, the data sets used in this analysis differ significantly from experiment to
 experiment, resulting in considerable dispersion in
the values of  ${\cal F}(1)\vert V_{cb}\vert$ and $\rho^2$ and hence in a
large $\chi^2$ of the combined fit,
summarized by HFAG~\cite{hfag07}:
\begin{eqnarray}
{\cal F}(1) \vert V_{cb} \vert &=&(35.89 \pm 0.56) \times 10^{-3}~,\\
\rho^2&=&1.23 \pm 0.05 \quad (\chi^2=37.8/17; {\rm CL}=0.026).
\nonumber
\end{eqnarray}
To convert this into a value of $\vert V_{cb} \vert$, we need to know 
${\cal F}(1)$. 
In terms of the perturbative (QED and QCD) and non-perturbative (leading $\delta_{1/m^2}$ and sub-leading
$\delta_{1/m^3}$) corrections, ${\cal F}(1)$ can be expressed as follows:
\begin{equation}
{\cal F}(1)= \eta_A\left[ 1 + \delta_{1/m^2} + \delta_{1/m^3} \right]~,
\end{equation}
where $\eta_A$ is the perturbative renormalization of the IW-function, known in the meanwhile to three
loops~\cite{Archambault:2004zs}. One- and two-loop corrections yield $\eta_A\simeq 0.933$
and the $O(\alpha_s^3)$
contribution amounts to $\eta_A^{(3)}=-0.005$.
 A Lattice-QCD calculation in the quenched approximation yields~\cite{Hashimoto:2001nb}
 $ {\cal F}(1)= 0.919^{+0.030}_{-0.035}$, which is now being reevaluated with
dynamical quarks. Taking into account this theoretical input, the value quoted
 at the Lepton-Photon-2007 Symposium is~\cite{Barberio-LP07}:
\begin{equation}
\vert V_{cb} \vert_{B \to D^* \ell \nu_\ell} = 
(39.1 \pm 0.65_{\rm exp} \pm 1.4_{\rm  theo})\times 10^{-3}.
\label{eq:vcb-excl}
\end{equation}
The resulting value of $\vert V_{cb} \vert $ is in excellent agreement with the
ones given in  (\ref{eq:vcb-bauer}) and (\ref{eq:vcb-babar}) obtained 
from the inclusive decays. 
 
\subsection{$\vert V_{ub} \vert$ from the decays $B \to X_u \ell \nu_\ell$}
\label{subsec:vubincl}

HQET techniques allow to calculate the inclusive decay rate $B \to X_u \ell \nu_\ell$ rather
accurately. However, the experimental problem in measuring this transition lies in the huge background from the dominant decays
$B \to X_c \ell \nu_\ell$ which can be brought under control only through severe cuts on the
kinematics. For example, these cuts are imposed on the lepton energy,
demanding $E_\ell > (m_B^2-m_D^2)/2m_B$, and/or 
the momentum transfer to the lepton pair $q^2$ restricting it below a threshold value $q^2 < q^2_{\rm max}$,
and/or the hadron mass recoiling against the leptons, which is required to satisfy
$m_X < m_D$. With these cuts,
the phase space of the decay $B \to X_u \ell \nu_\ell$ is  greatly reduced. A bigger problem is encountered
in the end-point region (also called the shape function region), where the leading power correction is no
longer $1/m_b^2$ but rather $1/m_b\Lambda_{\rm QCD}$, slowing the convergence of the expansion. 
Moreover, in the region of energetic leptons with low invariant mass hadronic states,
  the differential rate is sensitive
to the details of the shape function $f(k_+)$~\cite{Luke:2003nu}, where $ k_{+}=k_0+k_3$
with $k^\mu \sim O(\Lambda_{\rm QCD})$.

The need to know $f(k_+)$ can be circumvented to a large extent by doing a
 combined analysis of the data
on $B \to X_u \ell \nu_\ell$ and $B \to X_s \gamma$. Using 
the operator product expansion (OPE) to calculate the photon energy
spectrum in the inclusive decay $B \to X_s \gamma$,  
the leading terms in the spectrum (neglecting the bremsstrahlung corrections) can be re-summed into a
 shape function~\cite{shape}:
\begin{equation}
\frac{d\Gamma_s}{dx} = \frac{G_F^2 \alpha m_b^5}{32 \pi^4}\, |
          V_{ts} V_{tb}^*|^2\, |C_7^{\rm eff}|^2\, f(1-x) ~,
\end{equation}
where $x = \frac{2E_\gamma}{m_b}$, and $ C_7^{\rm eff}$ is 
an effective Wilson coefficient, characterizing the strength of the
electromagnetic dipole operator.  In the leading order, $E_\ell$- and $M_{X_u}$-spectra in
$B \to X_u \ell \nu_\ell$ are also governed by $f(x)$. Thus, 
$f(x)$ can be measured in $B \to X_s \gamma$ and used
in the analysis of data in $B \to X_u \ell \nu_\ell$.
  
 Following this argument,
 a useful relation emerges~\cite{Leibovich:1999xf,Neubert:2001sk,Bauer:2001mh}
\begin{equation}
\vert \frac{V_{ub}}{V_{tb}V_{ts}^*}\vert =\left( \frac{3 \alpha}{\pi} \vert C_7^{\rm eff}\vert^2
\frac{\Gamma_u(E_c)}{\Gamma_s(E_c)}\right)^{\frac{1}{2}}( 1 + \delta(E_c))~,
\label{eq:vubsgincl}
\end{equation}  
where 
\begin{eqnarray}
\Gamma_u(E_c) \equiv \int_{E_c}^{m_B/2} d E_\ell \frac{d\Gamma_u}{dE_\ell}~, \nonumber\\
& &\hspace*{-4.5cm}\Gamma_s(E_c) \equiv \frac{2}{m_b} \int_{E_c}^{m_B/2}
 d E_\gamma (E_\gamma -E_c)
 \frac{d\Gamma_s}{dE_\gamma}~,
\end{eqnarray}
and $\delta(E_c)$ incorporates the sub-leading terms in ${\cal O}(\Lambda_{\rm QCD}/m_b)$,
which can only be modeled at present.
 In addition, there are perturbative corrections to the
spectra and in the relation
 (\ref{eq:vubsgincl})~\cite{shape,Korchemsky:1994jb,Leibovich:1999xf}.

The other strategy is to extend the measurements of the inclusive decay
$B \to X_u \ell^+ \nu_\ell$ to the kinematic regions which are reachable by
the decay $B \to X_c \ell^+ \nu_\ell$, thus enlarging the region where the 
light-cone momentum component satisfies
 $P_+\equiv E_X - \vert \vec{P}_X\vert \gg
\Lambda_{\rm QCD}$, obviating the need to know the shape function. Both these
methods have been used in the determination of $\vert V_{ub} \vert$,
summarized below.

$\bullet$ Determination of $\vert V_{ub}\vert$ using the combined cuts on the
variables $m_X$ and $q^2$ following the suggestion by Bauer, Ligeti and Luke
(BLL)~\cite{Bauer:2001rc}. Taking the HQE parameter input from the analysis of
the decay $B \to X_c \ell \nu_\ell$ and $B \to X_s \gamma$, HFAG~\cite{hfag07} quotes an
average value $\vert V_{ub}\vert({\rm BLL})=(4.83 \pm 0.24 \pm 0.37) \times
10^{-3}$,
using the b-quark mass $m_b({\rm 1S})= (4.70 \pm 0.03)$ GeV in the so-called
$1S$-scheme.

$\bullet$ Determination of $\vert V_{ub}\vert$ from a fully differential decay
  rate for $B \to X_u \ell \nu_\ell$ based on the soft collinear effective theory
 (SCET) techniques~\cite{Bauer:2000ew,Bauer:2000yr,Bauer:2001yt,Beneke:2002ph},
 hereinafter called the Bosch-Lange-Neubert-Paz
  (BLNP) approach~\cite{Lange:2005yw}. The three independent kinematic variables 
   are chosen to be: $P_\ell =M_B - 2E_\ell$, $P_- =E_X + \vert
  \vec{P}_X\vert$, and $ P_+ =E_X - \vert\vec{P}_X\vert$, where $P_{\pm}$ are the
light-cone components of the hadronic final-state momentum along the jet
  direction, $E_X$ is the jet energy, $\vec{P}_X$ is the jet momentum, and
$E_\ell$ is the charged-lepton energy. In terms of these variables, the
  triple
differential distribution is:

\begin{eqnarray}
\frac{d^3\Gamma}{dP_+ dP_- dP_\ell} &=& \frac{G_F^2 \vert V_{ub}\vert^2}
{16\pi^2} (M_B -P_+)
\left\{(P_- + P_+)(M_B-P_- + P_\ell -P_+) {\it F}_1\right.\nonumber\\
& & \left. + (M_B-P_-) (P_- - P_+){\it F}_2 +(P_- - P_\ell)(P_\ell -P_+){\it F}_3\right\}.
\end{eqnarray} 
The ``structure functions'' ${\it F}_i$ can be expressed as  product of the
hard (perturbatively calculable) coefficient and a jet function, which are
convoluted with the soft light-cone distribution functions, the shape functions of the
$B$ meson. SCET allows to separate the two scales here, namely $\mu_h \sim m_b$
and $\mu_i \sim \sqrt{ m_b \Lambda_{\rm QCD}}$ and enables to sum large logarithms involving
the two scales $\mu_h$ and $\mu_i$. The dependence on the subleading shape
functions is studied by taking several models.  Fixing the HQE parameters (in the
so-called shape function scheme) to the values
$ m_b({\rm SF})=(4.63 \pm 0.06)$ GeV, $\mu_\pi^2({\rm SF})=(0.18 \pm 0.06)$ GeV$^2$, and the
exponential
form for the shape function, HFAG~\cite{hfag07} quotes  
$ \vert V_{ub} \vert = (4.31 \pm 0.17 \pm 0.35) \times 10^{-3}$ in the BLNP approach. 

$\bullet$ Determination of $\vert V_{ub}\vert$ using the so-called
Dressed Gluon Exponentiation (DGE) advocated by Andersen and
 Gardi~\cite{Andersen:2005mj}. The basic assumption of this approach is
that properly defined quark distribution in an on-shell heavy quark provides
a good approximation to the distribution in the meson. 
The problematic small hadronic mass $M_{X_u}$ region in the decay
 $B \to X_u  \ell^+ \nu_\ell$ is characterized by a large hierarchy in
the ratio $p_j^+/p_j^- \ll 1$ involving 
the partonic light-cone coordinates $p_j^+ \leq p_j^-\leq m_b$.
Defining the moments $n$ with respect to the powers of
$1 -p_j^+/p_j^-$, the region of small $p_j^+$ is probed by high moments
 $n \to \infty$, giving rise to the Sudakov logarithms $\ln n$.
 Infrared sensitivity appears in the moment-space
Sudakov exponents through infrared renormalons, leading to the divergence of
the (higher order perturbative) series in the Sudakov exponent. However,
 the leading infrared renormalon ambiguity cancels exactly against
the pole-mass renormalon ambiguity, which enters as a kinematic factor in the
differential decay width, making the on-shell perturbative
 calculation directly applicable for the phenomenology. This approach, applied to the
data  yields~\cite{hfag07} 
$\vert V_{ub}\vert =(4.34 \pm 0.16 \pm 0.25) \times 10^{-3}$ for the input
value $\overline{m_b}(\bar{m}_b)=(4.20 \pm 0.07)$ GeV.

Thus, the three theoretical approaches used in the determination of $\vert
V_{ub} \vert$ from the inclusive decays $B \to X_u \ell \nu_\ell$ give very
consistent values.

\subsection{$\vert V_{ub} \vert$ from exclusive decays}
 $\vert V_{ub} \vert$ has also been determined from the exclusive
decays $B \to \pi \ell \nu_\ell$. Theoretical accuracy is
limited by the imprecise knowledge of the form factors. 
A number of theoretical techniques has been used to determine them. These include,
among others,
Light-cone QCD sum rules~\cite{Ball:2004ye}, Quenched- and Unquenched-Lattice QCD
simulations~\cite{Dalgic:2006dt,Okamoto:2004xg,Abada:2000ty}. HFAG quotes the following
 extracted values of
$\vert V_{ub} \vert$ using the full $q^2$-range for the form factors
(in units of $10^{-3}$)~\cite{hfag07}:\\
$ |V_{ub}|= 3.43 \pm 0.10 ^{+0.67}_ {-0.42} $ ({\rm LCSR: Ball-Zwicky\cite{Ball:2004ye}});\\
$ |V_{ub}| = 3.17 \pm 0.10 ^{+ 0.77}_{-0.48} $ ({\rm Lattice: HPQCD\cite{Dalgic:2006dt}});\\ 
$ |V_{ub}| = 3.82 \pm 0.12 ^{+ 0.88}_{- 0.52}$ ({\rm Lattice: FNAL\cite{Okamoto:2004xg}}); \\
$ |V_{ub}| = 3.61 \pm 0.11  ^{+1.11}_{-0.57}$ ({\rm Lattice: APE\cite{Abada:2000ty}}).

These values of $\vert V_{ub} \vert$ from the exclusive decays (typically $3.5 \times
10^{-3}$) are smaller than the ones extracted from the inclusive measurements
(typically $4.3 \times 10^{-3}$). This mismatch is a mild irritant, waiting for  
 its resolution through progress in Lattice-QCD and more $B$-factory data.

To reduce the form-factor related uncertainties in extracting $\vert V_{ub} \vert$ from
exclusive decays $B \to (\pi,\rho) \ell \nu_\ell$, input from the rare $B$-decays
 $B \to (K,K^*) \ell^+\ell^-$ and HQET may be helpful.  A  proposal along these lines is
the so-called  Grinstein's double ratio which would
determine  $|V_{ub}|/|V_{tb} V_{ts}^*|$ from the end-point region of exclusive 
rare  $B$-meson decays~\cite{Grinstein:2004vb}. To carry out this program one has to measure
four distributions in the decays $B \to \rho \ell \nu_\ell$, $B \to K^* \ell^+\ell^-$, and $D \to (\rho,K^*) \ell \nu_\ell$.
With the help of this data and HQET, the ratio of the CKM factors  
$|V_{ub}|/|V_{tb} V_{ts}^*|$ can be determined through the double ratio 
\vspace*{-0.1cm}
\begin{equation}
\frac{\Gamma(\bar B \to \rho \ell \nu)}
     {\Gamma(\bar B \to K^* \ell^+ \ell^-)} \, 
\frac{\Gamma(D \to K^* \ell \nu)}
     {\Gamma(D \to \rho \ell \nu)}.
\end{equation}
At the $B$ factories, one expects enough data on these decays to allow a 10\% determination of $\vert V_{ub}\vert$
from  exclusive decays.
\section{Status of the Third Row of $V_{\rm CKM}$}
 FCNC  transitions $b \to s$ and $b \to d$ (as well as 
$s \to d$) give information on the third row of the CKM matrix $ V_{\rm CKM}$
and allow to search for physics beyond the SM. In the SM these
 transitions are generally dominated  
by the  (virtual) top quark contributions giving rise to the dependence 
on the matrix elements $\vert V_{tb}^*V_{ts}\vert$ 
(for $b \to s$ transitions) and 
$\vert V_{tb}^*V_{td}\vert$ (for $b \to d$ transitions). 
Of these, only the matrix element $\vert V_{tb}\vert$ has been 
measured by a tree process $t \to W b$ at
the Tevatron through the ratio

\begin{equation}
R_{tb}\equiv\frac{{\cal B}(t \to Wb)}{{\cal B}(t \to Wq)} =\frac{\vert 
V_{tb}\vert^2}{\vert V_{td}\vert^2 + \vert V_{ts}\vert^2 +
\vert V_{tb}\vert^2}\,.
\end{equation}

\noindent
The current measurements yield~\cite{Acosta:2005hr}: 
$R_{tb}=1.12^{+0.21}_{-0.19} ({\rm stat})^{+0.17}_{-0.13} ({syst})$,
yielding $R_{tb} > 0.61$ (at 95\% C.L.), which in turn gives
$\vert V_{tb} \vert > 0.78$ at 95\%~C.L..

Precision on the direct measurement of $\vert V_{tb} 
\vert$ is rather modest (unitarity gives $\vert V_{tb}\vert  \simeq 0.9992$.), 
which will be greatly improved, in particular, at 
a Linear Collider~\cite{AguilarSaavedra:2001rg}, but also at the LHC.
The corresponding measurements of $\vert V_{ts}\vert$ and $\vert V_{td}\vert$ from the 
tree processes are not on the cards. These matrix elements are determined by (loop)
induced processes discussed below. 

\subsection{Status of $\vert V_{td} \vert$}
The current best measurement of $\vert V_{td} \vert$ comes from 
$\Delta M_{B_d}$, the mass difference between the two mass eigenstates
of the $B_d^0$ - $\overline{B_d^0}$ complex. This has been measured in a 
number of experiments and is known to an accuracy of $\sim 1\%$; the current
world average is~\cite{hfag07} $\Delta M_{B_d}= 0.507 \pm 0.005$ (ps)$^{-1}$. 

In the SM, $\Delta M_{B_d}$ and its counterpart $\Delta M_{B_s}$, the mass 
difference in the $B_s^0$ - $\overline{B_s^0}$ system, are calculated by box 
diagrams, dominated by the 
$Wt$ loop. Since  $(M_W, m_t) \gg m_b$, $\Delta M_{B_d}$ is governed by the
short-distance physics. The expression for 
$\Delta M_{B_d}$ taking into account the perturbative-QCD corrections reads as 
follows~\cite{Buras:1984pq}
\begin{equation}
\Delta M_{B_d} = \frac{G_F^2}{6 \pi^2} \, \hat{\eta}_B \, 
\vert V_{td} V_{tb}^* \vert^2  \, 
M_{B_d}  \, (f_{B_d}^2\hat{B}_{B_d})  \, M_W^2  \, S_0(x_t)\,.
\label{buras1990fn}
\end{equation}
The quantity $\hat{\eta}_B$ is the next-to-leading log (NLL) perturbative QCD
 renormalization of the matrix element of the $(\vert \Delta 
B\vert =2, \Delta Q=0)$ four-quark operator, 
whose value is $\hat{\eta}_B=0.55 \pm 
0.01$ ~\cite{Buras:1990fn}; $x_t=m_t^2/M_W^2$ 
and $S_0(x_t)=x_tf_2(x_t)$ is an 
Inami-Lim function~\cite{Inami:1980fz}, with
\begin{equation}
f_2(x)=\frac{1}{4} +\frac{9}{4}\frac{1}{(1-x)} -\frac{3}{2}\frac{1}{(1-x)^2}
-\frac{3}{2}\frac{x^2 \ln x}{(1-x)^3}\,.
\label{inamilimf2}
\end{equation}
The quantity $f_{B_d}^2\hat{B}_{B_d}$ enters through the hadronic 
matrix element 
of the four-quark box operator,~defined~as:
\begin{equation}
\langle \bar{B}_q^0 | (\bar{b} 
\gamma_\mu(1-\gamma_5)q)^2 |B_q^0 \rangle \equiv \frac{8}{3}f_{B_q}^2 B_{B_q} M_{B_q}^2\,,
\label{bqhatfb}
\end{equation}
with $B_q =B_d$ or $B_s$. With $\Delta M_{B_d}$ and $\hat{\eta}_B$ known to a
high accuracy, and the current value of the top quark mass, defined in the
$\overline{\rm MS}$ scheme, $\bar{m}_t(m_t) =(162.3 \pm 2.2)$ GeV, leading to
$S_0(x_t)=2.29(5)$, i.e., $\delta S_0(x_t)/S_0(x_t)\simeq 
2.5\%$, the combined error from these sources is a few percent. This is 
completely negligible in comparison with the current theoretical uncertainty on the 
matrix element $f_{B_d}\sqrt{\hat{B}_{B_d}}$. For example, 
$O(\alpha_s)$-improved
calculations in the QCD sum rule approach yield
$f_{B_d}=(210\pm 19)$ MeV~\cite{Jamin:2001fw} and
 $f_{B_d}=(206\pm 20)$ MeV~\cite{Penin:2001ux},
whereas $\overline{B}_{B_d}$ in the $\overline{\rm MS}$ scheme in this approach 
is estimated as~\cite{Korner:2003zk}
$\overline{B}_{B_d} =1$ to within 10\%, yielding for the renormalization group 
invariant quantity $\hat{B}_{B_d} \simeq 1.46$, and an accuracy of 
about $\pm 15\%$ on $f_{B_d}\sqrt{\hat{B}_{B_d}}$.
Lattice calculations for $f_{B_d}\sqrt{\hat{B}_{B_d}}$ are uncertain due to the
chiral extrapolation. 
 Taking this into account, the current
unquenched lattice QCD calculations from the 
JLQCD and HPQCD Collaborations~yield~\cite{Aoki:2003xb,Gray:2005ad}
\begin{equation}
f_{B_d}\sqrt{\hat{B}_{B_d}}=(244 \pm 11 \pm 24)~{\rm MeV}\,,
\label{aoki2003xb}
\end{equation}
where the first error is statistical, and the second combines the uncertainty
from chiral extrapolation and other systematic errors. Using $\vert
V_{tb}\vert =1$, this yields~\cite{Okamoto:2005zg}
\begin{equation}
\vert V_{td} \vert = (7.4 \pm 0.8) \times 10^{-3}\,.
\label{vtdmass}
\end{equation}

\subsection{Present status of $\vert V_{ts} \vert$}
The quantity which currently provides the best determination of $\vert V_{ts} \vert$ 
is $\Delta M_{B_s}$, measured by D0 and CDF, with the CDF measurement being
more precise 
$\Delta M_{B_s}=(17.77 \pm 0.10 \pm 0.07)$ (ps)$^{-1}$~\cite{Abulencia:2006psa}.
The expression for $\Delta M_{B_s}$ in the SM can be obtained from the one for
$\Delta M_{B_d}$ (\ref{buras1990fn}) by the replacements: $M_{B_d} \to
M_{B_s}$, $(f_{B_d}^2\hat{B}_{B_d}) \to (f_{B_s}^2\hat{B}_{B_s})$, and, most importantly,
 $V_{td} \to V_{ts}$. Using the recent Lattice-QCD result
 $f_{B_s}\sqrt{\hat{B}_{B_s}}
=(281 \pm 21)$ MeV from the HPQCD collaboration~\cite{Dalgic:2006gp}
allows to determine $\vert V_{ts}\vert$ from $\Delta M_{B_s}$ with a precision of about 10\%,
comparable to the precision on $\vert V_{td}\vert$ in eq.~(\ref{vtdmass}).
 This is
not competitive with the indirect estimate of this matrix element from the
CKM unitarity, which yields $V_{ts} \simeq -V_{cb}= -4.1(1) \times 10^{-2}$.

The compatibility of the SM with the measured value of  $\Delta M_{B_s}$ is
usually tested by taking the value of $V_{ts}$ from the unitarity fits,
the measured value of $m_t$, and 
the Lattice-QCD value for $f_{B_s}\sqrt{\hat{B}_{B_s}}$. 
Following this reasoning, the Lattice-HPQCD collaboration
estimates  
$\Delta M_{B_s} ({\rm SM})=(20.3 \pm 3.0 \pm 0.8)$ (ps)$^{-1}$~\cite{Dalgic:2006gp}.
The corresponding estimate from the UTfit is
 $\Delta M_{B_s}({\rm SM})=(20.9 \pm 2.6)$  (ps)$^{-1}$~\cite{Bona:2007vi}
 and the CKMfitter yields
 $\Delta M_{B_s}({\rm SM})=(21. 7^{+5.9}_{-4.2})$  (ps)$^{-1}$~\cite{Charles:2004jd}.
 Thus, the experimental measurement of
$\Delta M_{B_s}$ is typically  about  $1\sigma$ below the SM estimates, with
 $\Delta M_s({\rm expt})/\Delta M_s({\rm SM})=0.88 \pm 0.13$~\cite{Dalgic:2006gp},
 $0.85 \pm 0.10$~\cite{Bona:2007vi} and $0.88 \pm 0.20$~\cite{Charles:2004jd}.

 Possible implications of the CDF measurement of $\Delta M_{B_s}$
  have been studied in several
 papers~\cite{Ciuchini:2006dx,Endo:2006dm,Ligeti:2006pm,Lenz:2006hd,Bona:2007vi}.
Perhaps, it is to the point to mention here that a value of $\Delta M_{B_s}
  \simeq 18$ (ps)$^{-1}$ was also hinted by the LEP data with the central
value being $17.7$ (ps)$^{-1}$. Anticipating this, the consequences
of an eventual measurement of $\Delta M_{B_s}$ around this value were worked
out some eight years ago~\cite{Ali:2000hy} for the parameters of the SM and
the minimal flavor-violating  supersymmetry~\cite{Ali:1999we}. It was
emphasized that a measurement
of  $\Delta M_{B_s}$ around this value would confirm the SM. The CDF
measurement of $\Delta M_{B_s}$ is very precise, and our knowledge of the CKM
parameters and the non-perturbative quantities has in the meanwhile also
improved, as discussed in this report. However, the bottom line remains the
same, namely that the SM has passed this crucial test comfortably. 
 
 The ratio of the mass 
differences $\Delta M_{B_d}/\Delta M_{B_s}$, now measured very precisely,
can be used to constrain the CKM ratio $\vert V_{td} 
\vert/\vert V_{ts} 
\vert$ using the SM relation~\cite{Ali:1978kk}:
\begin{equation}
\frac{\Delta M_{B_s}}{\Delta M_{B_d}}=\xi \, \frac{M_{B_s}}{M_{B_d}} \, 
\frac {\vert V_{tb}^*V_{ts}\vert^2}{\vert V_{tb}^*V_{td}\vert^2}\,,
\label{deltamsd}
\end{equation}
where $\xi \equiv f_{B_s}\sqrt{\hat{B}_{B_s}}/f_{B_d}\sqrt{\hat{B}_{B_d}}$.    
Theoretical  uncertainty in $\xi$ in the Lattice QCD approach
is arguably smaller compared to the one in $f_{B_s}\sqrt{\hat{B}_{B_s}}$, as in the 
SU(3) limit $\xi=1$, and the uncertainty is
actually in the  SU(3)-breaking corrections. 
Current estimate in 
the unquenched lattice calculations of $\xi$ is~\cite{Okamoto:2005zg} 
$\xi = 1.21 ^{+0.047}_{-0.035}$, which yields~\cite{Abulencia:2006psa}
\begin{equation}
 \vert V_{td}/V_{ts} \vert =0.2060 
\pm 0.0007 ({\rm exp}) ^{+0.008}_{-0.006} ({\rm th}).
\label{eq:vtdts-deltas}
\end{equation}
 This is by far the best measurement of this CKM ratio, and it provides a
 non-trivial constraint on the allowed profile of the unitarity triangle.
Combining eqs.~(\ref{eq:vtdts-deltas}) and  (\ref{vtdmass}) yields 
 $\vert V_{ts}\vert=(36 \pm 4)\times 10^{-3}$ with the error dominated by theory.
This completes our review of the CKM matrix elements $V_{ij}$.

\section{Radiative and Semileptonic Rare $B$ Decays}

 Two inclusive rare $B$-decays of current experimental
interest are $B \to X_s \gamma$  and $B \to X_s l^+ l^-$,
where $X_s $ is any charmless hadronic state with the strangeness quantum
number $s=1$.
They probe the SM  in the electroweak $b \to s$ penguin sector.
 The CKM-suppressed decays $B \to X_d \gamma$
and $B \to X_d l^+ l^-$ are difficult to measure due to low rates and 
formidable backgrounds. Instead, the search for $B \to X_d\gamma$ decay has been
carried out in the exclusive decay modes $B \to (\rho,\omega) \gamma$. Combined with the
decay $B \to K^* \gamma$, these decays provide  constraints on
the CKM parameters. The CKM-suppressed decays $B \to (\pi,\rho,\omega)
\ell^+\ell^-$ have not yet been measured.
 We review some of these rare $B$-decays in the context of the SM.

\subsection{$B \to X_s \gamma$: SM vs.~Experiments}
 The effective~Lagrangian~for the decays~$B \to X_s \gamma$ obtained by
 integrating
out the top quark and the heavy electroweak bosons 
reads as follows in the SM:
\begin{eqnarray}
 \vspace*{-0.3cm}
  {\cal L_{\rm eff}} \;&=& \;\;  
\; \frac{4 G_F}{\sqrt{2}} V_{ts}^* V_{tb} \sum_{i=1}^{8} C_i(\mu) Q_i~.
\label{eq:efflagrang}
\vspace*{-0.3cm}
\end{eqnarray}
In writing this, unitarity of the CKM matrix has been used and the term
proportional to the small matrix elements $V_{us}^* V_{ub}$ has been ignored.
 The complete list of operators and their Wilson coefficients in the NNLO approximation
evaluated at the scale $\mu=m_b$ can be seen elsewhere~\cite{Buchalla:1995vs}.

The dominant four-quark operators $Q_1$ and $Q_2$ are
\begin{eqnarray}
\label{eq:4-quark-operators}
 Q_1 = (\bar s \, c)_{V-A} \, (\bar c \, b)_{V-A}, 
& \qquad &
 Q_2 = (\bar s_i c_j)_{V-A} \,(\bar c_j b_i)_{V-A}, 
\end{eqnarray}
and the electromagnetic and chromomagnetic penguin 
operators $Q_7$ and $ Q_8$ are
\begin{eqnarray}
 Q_7 = -\frac{e \, \overline m_b(\mu)}{8\pi^2}\,
(\bar s \, \sigma^{\mu\nu} \, [1 + \gamma_5] \, b) 
 F_{\mu\nu} \, , 
& \quad &
 Q_8 = -\frac{g \, \overline m_b(\mu)}{8\pi^2}\,
(\bar s \,\sigma^{\mu\nu} \, [1 + \gamma_5] \, T^a \, b) 
 G^a_{\mu\nu} .\nonumber\\ 
\label{eq:penguin-operators} 
\end{eqnarray}
The factor $\overline {m_b}(\mu)$ is the $\overline{\rm MS}$ mass of the $b$ quark.
QCD-improved calculations in the effective theory require three steps (for a
review see~\cite{Hurth:2003vb}):\\
 (i) \underline{Matching} \, $C_i(\mu_0)$ ($\mu_0 \sim M_W, m_t$):
They have been calculated up to three loops~\cite{Bobek:199mk,Misiak:2004ew}.
 The three-loop matching is found to have  
less than $2\%$ effect on ${\cal B}(B \to X_s \gamma)$~\cite{Misiak:2004ew}.\\
 (ii) \underline{Operator mixing:} This involves calculation of the anomalous dimension matrix,
which is expanded in $\alpha_s(\mu)$.
 The anomalous dimensions up to  $\alpha_s^2(\mu)$ 
are known~\cite{Chetyrkin:1996vx} since a decade, and  the $\alpha_s^3(\mu)$ calculations 
have been completed recently in a series of papers~\cite{Gorbahn:2004my,Gorbahn:2005sa,Czakon:2006ss}.

\noindent
 (iii) \underline{Matrix elements} $\langle O_i \rangle (\mu_b)$ ($\mu_b \sim m_b$):
The first two terms in the expansion in
$\alpha_s(\mu_b)$ are known since long~\cite{Greub:1996tg}.
Exact results to ${\cal O}(\alpha_s^2)$  were obtained for 
$Q_7$ in~\cite{Blokland:2005uk,Asatrian:2006ph} and for 
$Q_8$ in~\cite{greub_prep}.
For $Q_1$ the virtual corrections at  ${\cal O}(\alpha_s)$ were 
calculated in~\cite{Greub:1996tg,Buras:2001mq,Buras:2002tp}, 
but those at ${\cal O}(\alpha_s^2)$
are known only in the large-$\beta_0$ 
limit~\cite{Bieri:2003ue}. A calculation that goes beyond this
approximation by employing an interpolation in the charm quark mass $m_c$
was reported in ~\cite{Misiak:2006ab}, and has  been used in estimating
the NNLO branching fraction for the inclusive decay
 $B \to X_s \gamma$~\cite{Misiak:2006zs}.
 Finally, one has to add the
Bremsstrahlung contribution $b \to s \gamma g$ to get the complete decay rate,
which in $O(\alpha_s)$ was done in~\cite{Ali:1990tj} and in $O(\alpha_s^2)$ in
\cite{Melnikov:2005bx}.

In the $\overline{\rm MS}$ scheme, the NNLO branching ratio for $E_\gamma >
 1.6$ GeV 
 is calculated as~\cite{Misiak:2006zs}:
\begin{equation}
{\cal B} (B \to X_s \gamma)_{\rm SM}= (3.15 \pm 0.23) \times 10^{-4}~.
\label{eq:misiak96}
\end{equation}
This amounts to a theoretical precision
of about $8\%$, comparable to the current experimental precision~\cite{hfag07} 
\begin{equation}
{\cal B} (B \to X_s \gamma)_{\rm Expt.}= (3.55  \pm 0.24 ^{+0.09}_{-0.10}\pm 0.03) \times 10^{-4}~.
\label{eq:bsghfag04}
\end{equation}
A comparison of the two shows that the SM estimate is in agreement with data
though the SM central value lies below the experiment by about $1\sigma$.
This allows for speculations about a beyond-the-SM contribution
interfering constructively with the SM amplitude. A case in point is a 2Higgs
doublet model (2HDM); the preferred value is $m_{H^+}\simeq 650$ GeV with a
95\% C.L. lower bound $m_{H^+} > 295$ GeV. However, more conservatively, the
proximity of ${\cal B}(B \to X_s \gamma)$ in the SM and experiment puts bounds
on the parameters characterizing new physics. This has been worked out,
together with other constraints, in the context of
 supersymmetry~\cite{Ellis:2007fu}.

 The current (NNLO) theoretical precision on ${\cal B} (B \to X_s \gamma)$
has also been investigated  in the context of SCET using a multi-scale OPE
involving three low energy scales: $m_b$, $\sqrt{m_b \Delta}$ and $\Delta=m_b-2E_0$, where 
$E_0$ is the lower cut on the photon energy. Large logarithms associated with
these scales are summed at NLL order. The sensitivity to the scale $\Delta$
 introduces additional uncertainties. Taking this into account, 
Becher and Neubert~\cite{Becher:2006pu} estimate ${\cal B}(B \to X_s \gamma) =(2.98 \pm 0.26)
\times 10^{-4}$, which increases the departure of the SM from data to about $1.4\sigma$.

\subsection{$B \to X_s  \ell^+ \ell^-$: SM vs.~Experiments}
To study the decays $B \to X_s \ell^+ \ell^-$, one has to extend the operator
basis in
the effective Lagrangian (\ref{eq:efflagrang}) by adding two semileptonic
operators~\cite{Grinstein:1988me}:
\begin{eqnarray}
\label{eq:4-semilept-operators}
 Q_9 = \frac{e^2}{16\pi^2} (\bar s \, \gamma_\mu \, b)_{V-A} \,
 (\bar \ell \, \gamma^\mu \, \ell)~, 
& \qquad &
  Q_{10} = \frac{e^2}{16\pi^2}(\bar s \, \gamma_\mu \, b)_{V-A} \,
 (\bar \ell \, \gamma^\mu
  \gamma_5 \, \ell)~.\nonumber
\end{eqnarray}
 The corresponding Wilson coefficients $C_9(\mu)$ and $C_{10}(\mu)$ have the
following perturbative expansion:

\begin{eqnarray}
\label{eq:C910-expan}
C_{9} &=& \frac{4\pi}{\alpha_s(\mu)} C_9^{(-1)}(\mu) + C_9^{(0)}(\mu) +
       \frac{\alpha_s(\mu)}{4\pi} C_9^{(1)} (\mu) + ...\nonumber\\
& & C_{10} = C_{10}^{(0)} +  \frac{\alpha_s(M_W)}{4\pi} C_{10}^{(1)} +...~.
\end{eqnarray}
After an expansion in $\alpha_s$, the term $C_9^{(-1)}(\mu)$ reproduces the
dominant part of the electroweak logarithms that originate from photonic penguins
with charm quark loops: 
\begin{equation}
\frac{4\pi}{\alpha_s(m_b)} C_9^{(-1)}(m_b) =\frac{4}{9} \ln
\frac{M_W^2}{m_b^2} + O(\alpha_s)~,
\end{equation}
leading to $\frac{4\pi}{\alpha_s(m_b)} C_9^{(-1)}(m_b) \simeq 2 $. With
$C_9^{(0)}(m_b) \simeq 2.2$, one needs to calculate in the NNLO accuracy.
The NNLO calculation of the decay   $B \to X_s l^+ l^-$ corresponds to the NLO
calculation of $B \to X_s \gamma$, as far as the
number of loops in the diagrams is concerned.  

The process $B \to X_s \ell^+\ell^-$ differs greatly from the radiative decay $B \to
X_s \gamma$ as far as non-perturbative contributions are concerned. The
largest effect in $B \to X_s \ell^+\ell^-$  from the intermediate
$c\bar{c}$ states  comes from the
resonances $J/\psi$, $\psi^\prime$ and $\psi^{\prime\prime}$ decaying to
$\ell^+\ell^-$, which can be either modeled, for example, as done by Kr\"uger and
Sehgal~\cite{Kruger:1996cv}
using dispersion relations and data on $\sigma_(e^+e^- \to c\bar{c}\to {\rm
  hadrons})$, or else experimental cuts are imposed on $q^2$ to remove the resonant
regions and the short-distance contribution is extrapolated through these
cuts. Then, there are factorizable $1/m_c$ and $1/m_b$ power corrections, similar to those
in $B \to X_s \gamma$, which can be calculated using the OPE and HQET. As is
the case for $B \to X_s \gamma$ and $B \to X_u \ell
\nu_\ell$, there are no $1/m_b$ corrections. The $O(1/m_b^2)$ corrections 
in this framework were calculated first in~\cite{Falk:1993dh} and corrected in
~\cite{Ali:1996bm}. The $O(1/m_b^3)$ corrections were calculated in ~\cite{Bauer:1999kf}.
 The $1/m_c$ factorizable power corrections were calculated in~\cite{Chen:1997dj}.

 Including the leading power corrections in $1/m_b$  and
 $1/m_c$
and taking into account various parametric uncertainties, the branching ratios for the decays
$B \to X_s \ell^+ \ell^-$ in NNLO are~\cite{Ali:2002jg}:
\begin{eqnarray}
{\cal B}(B \to X_s e^+ e^-)_{\rm SM}  &\simeq& {\cal B}(B \to X_s \mu^+ \mu^-)_{\rm SM}
= (4.2 \pm 0.7) \times 10^{-6}~,
\label{eq:aghl}
\end{eqnarray}
where a dilepton invariant mass cut, $m_{\ell \ell} > 0.2 $ GeV, has been assumed for
comparison with data given below. These estimates make use of the NNLO calculation by
 Asatryan et al.~\cite{Asatryan:2001zw}, 
restricted to $\hat s \equiv q^2/m_b^2< 0.25$. The spectrum for 
 $\hat s> 0.25$ has been obtained from the NLO calculations using the scale $\mu_b \simeq m_b/2$,
as this choice of the scale reduces the NNLO contributions. Subsequent NNLO calculations
covered the entire dilepton mass spectrum and are numerically in agreement with this
procedure, yielding
${\cal B}(B \to X_s \mu^+ \mu^-)_{\rm SM}= (4.6 \pm 0.8)
 \times 10^{-6}$~\cite{Ghinculov:2003qd,Bobeth:2003at} .
 The difference in the central
values in these results and (\ref{eq:aghl}) is of parametric origin.

 The BABAR and BELLE collaborations have measured the
invariant dilepton and hadron mass spectra in $B \to X_s \ell^+\ell^-$. Using
the SM-based calculations
 to extrapolate
through the cut-regions, the current averages of the branching ratios are~\cite{hfag07}:
\begin{eqnarray}
{\cal B}(B \to X_s e^+ e^-)&=&~(4.7 \pm 1.3) \times 10^{-6},
\nonumber\\   
{\cal B}(B \to X_s \mu^+ \mu^-)&=& ~(4.3 ^{+1.3}_{-1.2})\times 10^{-6}~,
\nonumber\\ 
&&\hspace*{-3.2cm}{\cal B}(B \to X_s \ell^+ \ell^-)= ~(4.5 ^{+1.03}_{-1.01})\times 10^{-6}.
\label{eq:bsll-expt}
\end{eqnarray}
Thus, within the current experimental accuracy, which is typically 25\%, data and the SM
agree with each other in the $b \to s$ electroweak penguins. The low
$q^2$-region
(say, $q^2 < 8$ GeV$^2$), which allows the most precise comparison with the SM,
suffers both from the statistics and a cut on the invariant hadronic mass
recoiling against the dilepton. A  cut $m_X > 2 GeV$ and $m_X > 1,8$
GeV
have been used by the BELLE and BABAR collaborations, respectively. The
effects of these cuts
have been studied in the Fermi-motion model~\cite{Ali:1998ku}, which has been
used
in the experimental analysis of the data so far. Subsequently, the $B \to X_s
\ell^+\ell^-$
rate with an $m_X$ cut in the low-$q^2$ region has been calculated using the
$B \to X_s \gamma$ shape function~\cite{Lee:2005pwa}. This work, whose impact
on the analysis of the $B \to X_s \ell^+\ell^-$ data has yet to be studied,
reduces some of the theoretical errors in the SM estimates given in
eq.~(\ref{eq:aghl}). In the same vein, it has also been recently
  argued~\cite{Ligeti:2007sn} that
the non-perturbative uncertainties in the large-$q^2$ region $( q^2 \geq
14~{\rm GeV}^2)$ can be
significantly reduced by normalizing the partial decay width of $B \to X_s \ell^+\ell^-$ 
with the corresponding partial width of the decay  
$ B \to X_u \ell \nu_\ell$. With more data from the $B$ factories, these
theoretical
developments will enable a more precise test of the SM in the
$B \to X_s \ell^+ \ell^-$ decays.

 The measurements (\ref{eq:bsghfag04}) and
(\ref{eq:bsll-expt}) provide valuable constraints
 on beyond-the-SM physics scenarios.
 Following the earlier analysis to determine the Wilson coefficients  in $b \to s$
 transitions~\cite{Ali:1994bf,Ali:2002jg,Hiller:2003js},
it has been recently argued~\cite{Gambino:2004mv} that data now disfavor solutions in which the
 coefficient $C_7^{\rm eff}$ is similar in magnitude but opposite in sign to the SM coefficient.

Exclusive decays $B \to (K,K^*)\ell^+\ell^-$ $(\ell^\pm=e^\pm,\mu^\pm)$ have also been measured by the
BABAR and BELLE collaborations, and the current world averages of the branching ratios are~\cite{hfag07}:
\begin{eqnarray}
{\cal B}(B \to K \ell^+\ell^-)&=&~(3.9 \pm 0.6) \times 10^{-7},
\nonumber\\   
{\cal B}(B \to K^* e^+ e^-)&=& ~(11.3 ^{+2.8}_{-2.6})\times 10^{-7}~,
\nonumber\\ 
&&\hspace*{-3.2cm}{\cal B}(B \to K^* \mu^+ \mu^-)= ~(10.3
^{+2.6}_{-2.3})\times 10^{-7}~,\\
&&\hspace*{-3.2cm}{\cal B}(B \to K^* \ell^+ \ell^-)= ~(9.4
^{+1.7}_{-1.6})\times 10^{-7}~.
\label{eq:bkstll-expt}
\end{eqnarray}
They are also in agreement with the SM-based estimates of the same.
A calculation based on the light cone QCD sum rules for the form factors~\cite{Ali:1991is}
yields~\cite{Ali:2002jg}:
${\cal B}(B \to K \ell^+ \ell^-)= (3.5 \pm 1.2)\times 10^{-7}$, 
${\cal B}(B \to K^* e^+ e^-)= (15.8 \pm 4.9)\times 10^{-7}$,
and  ${\cal B}(B \to K^* \mu^+ \mu^-)= (11.9 \pm 3.9)\times 10^{-7}$ with the errors
dominated by uncertainties on the form factors. In the future, these errors
can be reduced by using the data on $B \to (\pi, \rho) \ell \nu_\ell$
to determine the $B \to (\pi,\rho)$ form factors. This information can be combined with
estimates of the SU(3)-symmetry breaking to determine the  $B \to (K, K^*)$ form factors,
enabling  to predict the FCNC decay rates and spectra more
precisely. For the low invariant mass of the dileptons, say $q^2 < 8$ GeV$^2$,
the SCET framework can be employed to reduce the number of form factors and improve the
perturbative aspects of these decays.

 The Forward-Backward (FB) asymmetry in the
decay $B \to X_s \ell^+ \ell^-$, defined as~\cite{Ali:1991is} 
\begin{eqnarray}
\bar {\cal A}_{\rm FB}(q^2) &=& \frac{1}{d {\cal B} ( B\to X_s \ell^+\ell^-) /d q^2  }
  \int_{-1}^1 d\cos\theta_\ell ~
 \frac{d^2 {\cal B} ( B\to X_s \ell^+\ell^-)}{d q^2  ~ d\cos\theta_\ell}
\mbox{sgn}(\cos\theta_\ell)~,\nonumber
\vspace*{-1.5cm} 
\end{eqnarray}
provides additional constraints on the Wilson coefficients. In particular, 
 the location of the zero-point of this asymmetry (called below
$q_0^2$) is a precision tests of the SM. In NNLO, one has the following
 predictions
for the inclusive decays $B \to X_s \ell^+\ell^-$:
 $q^2_0 =   (3.90 \pm 0.25)~{\rm GeV^2}~[(3.76 \pm 0.22_{\rm theory} \pm 0.24_{m_b})~{\rm GeV^2 }]$,
obtained by Ghinculov et al.~\cite{Ghinculov:2002pe} [Asatrian et
 al.~\cite{Asatrian:2002va}].

 In the SM (and its extensions in which the operator basis remains unchanged),
the  FB-asymmetry in $B \to K \ell^+ \ell^-$ is zero and in $B \to K^* \ell^+ \ell^-$
it  depends on the decay form factors. 
 Model-dependent studies yield small form factor-related
uncertainties in the zero-point of the asymmetry
 $\hat{s}_0=q_0^2/m_B^2$~\cite{Burdman:1998mk}.
 HQET provides a symmetry argument why the uncertainty in
$\hat{s}_0$ can be expected to be small which is determined by~\cite{Ali:1999mm} 
$ C_9^{eff}(\hat{s}_0) = -\frac{2 m_b}{M_B \hat{s}_0} C_7^{eff}$.
 However, 
$O(\alpha_s)$ corrections to the HQET-symmetry relations
lead to substantial change in the profile of the FB-asymmetry
function as well as a significant shift 
in $\hat{s}_0$~\cite{Beneke:2000wa,Beneke:2001at}. They have been
worked out for  $B \to K^* \ell^+\ell^-$  using
 SCET~\cite{Ali:2006ew}. 
Restricting ourselves to the kinematic region where the light $K^\ast$
 meson moves fast and can be viewed approximately as
a collinear particle, a
factorization formula for the decay amplitude of $B \to K^* \ell^+\ell^-$,
to leading power in $1/m_b$, has been derived in SCET~\cite{Ali:2006ew}.
This coincides {\it formally} with the formula obtained  earlier 
by Beneke et al.~\cite{Beneke:2000wa}, using the QCD factorization
approach~\cite{Beneke:1999br,Beneke:2001ev}, 
but is valid to all orders of $\alpha_s$:
\begin{eqnarray}
\langle K_a^\ast \ell^+ \ell^- \vert H_{eff} \vert B \rangle &=& T^I_a(q^2)
\zeta_a(q^2) +
\label{eq:SCETform} \\
& + & 
\sum_{\pm} \int_0^\infty \frac{d\omega}{\omega}
\phi^{B}_{\pm}(\omega) 
\int_0^1 du ~\phi_{K^\ast}^{ a}(u)T^{II}_{a,\pm}
(\omega, u,q^2)~,\nonumber
\end{eqnarray}
where $a=\parallel,\perp$ denotes the polarization of the $K^\ast$
meson. The functions $T^I_a$ and $T^{II}_{a,\pm}$ are perturbatively
calculable. $\zeta_a(q^2)$ are the soft form factors defined in SCET while
$\phi^{B}_{\pm}(\omega)$ and $\phi_{K^\ast}^{a}(u)$ are the light-cone distribution
amplitudes (LCDAs) for the B and $K^\ast$ mesons, respectively.
In particular, the
location of the zero of the forward-backward asymmetry in $B \to K^* \ell^+\ell^-$,
 $q_0^2$, can be
predicted more precisely in SCET due to the improved theoretical precision on the
scale dependence of $q_0^2$. 

Including the order $\alpha_s$ corrections, the analysis in \cite{Ali:2006ew}
estimates the
zero-point of the FB asymmetry to be
\begin{equation}
q^2_0=(4.07^{+0.16}_{-0.13})~ \mbox{GeV}^2~,
\end{equation}
of which the scale-related uncertainty is
 $\Delta(q_0^2)_{\rm scale}=^{+0.08}_{-0.05}$
 GeV$^2$ for the range
 $m_b/2 \leq \mu_h \leq 2 m_b$ together with the jet function scale
$\mu_l=\sqrt{\mu_h \times 0.5~\mbox{GeV}}$. This is to be compared with
the result given in~\cite{Beneke:2001at}, also obtained in the
absence of $1/m_b$ corrections:
$ q^2_0=(4.39^{+0.38}_{-0.35})~ \mbox{GeV}^2$. Of this the largest single
uncertainty (about $\pm 0.25~ \mbox{GeV}^2 $) is attributed to the scale
dependence.  The difference in the
estimates of the scale dependence of $q_0^2$  in~\cite{Ali:2006ew} and
\cite{Beneke:2001at} is both due to the incorporation of the SCET
 logarithmic resummation (done in~\cite{Ali:2006ew}) 
 and the different (scheme-dependent)
definitions of the effective form factors for the SCET currents used in these
references.
Power corrections in $1/m_b$  are probably
comparable to the $O(\alpha_s)$ corrections, as argued in~\cite{Beneke:2001at}.
So far, $q_0^2$ has not been measured experimentally. 

BELLE has published the first measurements~\cite{Nakao-LP07,Ishikawa:2006fh}
 of the
forward-backward asymmetry (FBA)~\cite{Ali:1991is}. The best-fit
results by BELLE for the Wilson coefficient ratios for negative value of $C_7$,
$\frac{C_9}{C_7} = -15.3 ^{+3.4}_{-4.8} \pm 1.1$ and
$\frac{C_{10}}{C_7} = 10.3 ^{+5.2}_{-3.5} \pm 1.8$,
are consistent with the SM values $C_9/C_7 \simeq -13.7$ and
$C_{10}/C_{7} \simeq +14.9$, evaluated in the NLO approximation.
However, for the positive value of $C_7$, the measurements lead to
$\frac{C_9}{C_7} = -16.3 ^{+3.7}_{-3.7} \pm 1.1$ and
$\frac{C_{10}}{C_7} = +11.1 ^{+6.0}_{-3.9} \pm 1.8$ and the two solutions
are of comparable significance.
With more data at the current B factories, and yet more anticipated
at the LHC,  these measurements are expected to become very
 precise, providing a precision test of the SM in the flavor sector.

\vspace*{-0.3cm}
\subsection{$B \to V \gamma $: SM vs.~Experiments}
The decays $B \to V \gamma $ $(V= K^*,\rho,\omega $) have been calculated in the 
 NLO approximation using the effective Lagrangian given in (\ref{eq:efflagrang}) and its analogue for
 $b \to d$ transitions. Two dynamical approaches, namely the QCD
 Factorization~\cite{Beneke:1999br}
and  pQCD~\cite{Keum:2000ph}
 have been employed to establish factorization of the radiative decay
 amplitudes in the heavy-quark limit.
 We illustrate the QCD-F method, where this factorization is worked out
for the $B \to V \gamma$
~\cite{Ali:2001ez,Bosch:2001gv,Beneke:2001at,Kagan:2001zk,Ali:2004hn,Bosch:2004nd,Beneke:2004dp}
 (see \cite{Ali:2006fa,Ball:2006nr}
for phenomenological updates in NLO, and ~\cite{Keum:2004is,Lu:2005yz}
for the alternative ``perturbative QCD'' approach). In particular,
the matrix element of a given operator in the effective weak
Hamiltonian  can be written in the form
\begin{equation} 
\left \langle V \gamma \left | Q_i 
\right | \bar B \right \rangle = 
F^{B \to V_\perp}  \, T_i^{\rm I} + 
\int d\omega \, du \,
\phi^B_+ (\omega) \, \phi^V_\perp (u) \, 
T^{\rm II}_i (\omega,u) \, .
\label{eq:ff} 
\end{equation}
The non-perturbative effects are contained 
in $F^{B \to V_\perp}$, the $B \to V$ transition 
form factor at $q^2 = 0$, and 
in $\phi^B_+$ and $\phi^V_\perp$, 
the leading-twist LCDAs of the $B$- and $V$-mesons. The hard-scattering 
kernels~$T^{\rm I}_i$ and~$T^{\rm II}_i$ include 
only short-distance effects and are calculable in 
perturbation theory. Contributions to the kernel
$T^{\rm I}$ are closely related to the virtual corrections 
to the inclusive decay rate, and are referred to as vertex corrections.  
Those to the kernel $T^{\rm II}$ are
related to parton exchange with the light quark in the 
$B$-meson, a mechanism  commonly referred to as 
hard spectator scattering.  It is expected that the factorization 
formula is  valid up to corrections of ${\cal O}(\Lambda_{\rm QCD}/m_b)$. 

The derivation of the factorization formula from a two-step
matching procedure in SCET has provided  additional insight into its structure.   
The technical details for $B\to V\gamma$ in NLO have been provided
in \cite{Chay:2003kb,Becher:2005fg}.  In the SCET 
approach the factorization formula is written as 
\begin{equation}
\label{eq:SCETff}
\left \langle V \gamma \left |  Q_i 
\right | \bar B \right \rangle  = \Delta_i C^{A}\zeta_{V_\perp}  + 
\frac{\sqrt{m_B}F f_{V_\perp}}{4}\int d\omega \, du \,
\phi^B_+ (\omega) \, \phi^V_\perp (u) \, 
t^{\rm II}_{i} (\omega,u)\,,
\end{equation}
where $F$ and $f_{V_\perp}$ are meson decay constants.
The SCET form factor $\zeta_{V_\perp}$ is related to the QCD form 
factor through perturbative and power corrections
\cite{Beneke:2000wa,Beneke:2003pa,Lange:2003pk,Beneke:2004rc,
Beneke:2005gs,Hill:2004if,Becher:2004kk}.  
In SCET the perturbative hard-scattering kernels are the 
matching coefficients $\Delta_i C^A$ and 
$t_i^{\rm II}$.  They are known completely to next-to-leading order
(NLO) $({\cal O}(\alpha_s))$ in renormalization-group (RG) 
improved perturbation theory \cite{Becher:2005fg}. 
Recently,  important steps towards a complete analysis
at the next-to-next-to-leading order (NNLO) in $B \to V \gamma$ decays have
been derived in~\cite{Ali:2007sj} by obtaining 
full results for the hard-scattering kernels
for the dipole operators $Q_7$ and $Q_8$, and partial results 
for $Q_1$, valid in the large-$\beta_0$ limit and neglecting
NNLO corrections from spectator scattering. In addition, this work provides
the virtual corrections to this order for the $B \to V \gamma$ decays, as
they can not be obtained from the published calculations for the
inclusive decay $B \to X_s \gamma$, discussed previously.

In SCET the hard-scattering kernel $t_i^{\rm II}$ for a given
operator is sub-factorized into the convolution of a hard-coefficient 
function with a universal jet function, in the form
\begin{equation}\label{eq:CJ}
t_i^{{ \rm II}}(u,\omega)=\int_0^1 d\tau \Delta_i C^{B1}(\tau) 
j_\perp(\tau,u,\omega)
\equiv  \Delta_i C^{B1}\star j_\perp.
\end{equation}
The hard coefficients $\Delta_i C^{B1}$ contain 
physics at the hard scale $m_b$, while 
the jet function $j_\perp$ contains physics at the hard-collinear 
scale $\sqrt{m_b \Lambda}$.  The hard coefficient is 
identified in a first step of matching ${\rm QCD} \to{\rm SCET}_{\rm I}$, 
and the jet function in a second step of matching 
${\rm SCET}_{\rm I}\to {\rm SCET}_{\rm II}$. Details in NLO have been 
worked out for  $B\to V\gamma$ in ~\cite{Chay:2003kb,Becher:2005fg}.

The effective field-theory techniques are crucial for
providing a field-theoretical definition of the objects 
in (\ref{eq:SCETff}),  and for resumming large perturbative
logarithms of the ratio $m_b/\Lambda_{\rm QCD}$ in  $t_i^{\rm II}$.  
In the effective-theory approach resummation is carried out
by solving the renormalization-group equations for the matching
coefficients $\Delta_i C^{B1}$. Since these coefficients
enter the factorization formula in a convolution with the jet
function $j_\perp$, their anomalous dimension is a distribution 
in the variables $\tau$ and $u$. The evolution equations must be solved 
before performing the convolution with $j_\perp$.  Therefore,
resummation is not possible in the original QCD 
factorization formula (\ref{eq:ff}), where 
the hard-scattering kernels $T_i^{\rm II}$ are obtained
only after this convolution has been  carried out.

Using the SCET framework, the 
 branching ratios in the NNLO are as follows~\cite{Ali:2007sj}:
\begin{eqnarray}
{\cal B}_{\rm NNLO} (B^0 \to K^{*0} \gamma) & \simeq & 
(4.3 \pm 1.4)\times 10^{-5} \,,
\nonumber \\[-1mm]
{\cal B}_{\rm NNLO} (B^\pm \to K^{*\pm} \gamma) & \simeq &
 (4.6 \pm 1.4) \times 10^{-5} \,,
\nonumber \\[-1mm]
{\cal B}_{\rm NNLO} (B_s^0 \to \phi \gamma) & \simeq & 
(4.3 \pm 1.4)\times 10^{-5}~.
\nonumber
\end{eqnarray}
It should be noted that, very much like the $B \to X_s \gamma$ case, the
complete NNLO calculations for the virtual corrections to the matrix element of
the operator $O_1$ in $B \to K^* \gamma$ are not yet at hand.
In addition, the hard spectator corrections from this operator are calculated
only in NLO. In the NNLO branching
ratios quoted above, the errors are increased to take these missing pieces into account.

The above theoretical branching ratios, compared with the current experimental
measurements~\cite{Wicht:2007rt,hfag07}, yield the following
 results~\cite{Ali:2007sj}: 
\begin{eqnarray}
\frac{{\cal B}_{\rm NNLO}(B^0 \to K^{*0} \gamma) }
{{\cal B}_{\rm Expt}(B^0 \to K^{*0} \gamma)} &=& 1.1 \pm 0.35 \pm 0.06,
\nonumber\\ 
\frac{{\cal B}_{\rm NNLO}(B^\pm \to K^{*\pm} \gamma) }
{{\cal B}_{\rm Expt}(B^\pm \to K^{*\pm} \gamma)} &=& 1.1 \pm 0.35 \pm 0.07,
\nonumber\\
\frac{{\cal B}_{\rm NNLO}(B_s^0 \to \phi \gamma) }
{{\cal B}_{\rm Expt}(B_s^0 \to \phi \gamma)} &=& 0.8 \pm 0.2 \pm 0.3~.
\nonumber
\label{eq:bkstar-exp}
\end{eqnarray}

The decays $B \to (\rho,\omega) \gamma$ involve in addition to the (short-distance) penguin
amplitude also significant long-distance
contributions, in particular in the decays $B^\pm \to \rho^\pm \gamma$.
 In the factorization approximation,  typical Annihilation-to-Penguin
amplitude ratio is estimated as~\cite{Ali:1995uy}:  
$\epsilon_{\rm A}(\rho^\pm \gamma)= 0.30 \pm 0.07$.
$O(\alpha_s)$ corrections to the annihilation amplitude in
$B \to \rho \gamma$ are not known; also the proof of factorization of
this amplitude is still not at hand.  
 The annihilation contribution to the decays
$B^0 \to \rho^0 \gamma$ and $B^0 \to \omega \gamma$  is expected to be suppressed 
(relative to the corresponding amplitude in $B^\pm \to \rho^\pm \gamma$) due to   
the electric charges ($Q_d/Q_u=-1/2$) and the  color factors, and the 
corresponding $A/P$ ratio for these decays is estimated as
$\epsilon_{\rm A}(\rho^0 \gamma) \simeq - \epsilon_{\rm A}(\omega \gamma)
 \simeq 0.05$.

The decay amplitudes for $B \to (\rho,\omega) \gamma$ depend on the CKM matrix
elements $V_{td}^* V_{tb}$ (from the penguin diagrams) and $V_{ub}^*
V_{ud}$ (from the  annihilation diagrams). Hence, these decays 
provide potentially very powerful constraints on the CKM parameters, $\bar
\rho$ and $\bar \eta$. Since a large number of observables can be measured
in these decays, such as the individual branching ratios for $B^\pm \to
\rho^\pm \gamma$ and $B^0 \to (\rho^0, \omega) \gamma$, isospin- and
SU(3)-violating asymmetries in the decay rates, and direct and time-dependent CP asymmetries,
they have been studied theoretically in a number of
 papers~\cite{Ali:2001ez,Bosch:2001gv,Ali:2004hn,Bosch:2004nd,Ali:2006fa,Ball:2006nr,Li:2007bh}.
Experimentally, a beginning has been made in the measurements of the 
$b \to d \gamma$ transition through the measurements of the branching
ratios for $B \to (\rho,\omega) \gamma$, reported by BABAR and BELLE. Current
measurements are not very precise, as can be seen from the current world
 averages~\cite{hfag07} (in units of $10^{-6}$): 
 ${\cal B} (B^\pm \to \rho^\pm \gamma)=0.96\pm 0.23$, 
${\cal B} (B^0 \to \rho^0 \gamma)=0.77\pm 0.14$ and
 ${\cal B} (B^0 \to \omega \gamma)=0.41\pm 0.15$.  In addition, theoretical
 estimates
suffer from large hadronic uncertainties, dominated by the imprecise knowledge
of the form factors. Hence, the resulting constraints on the CKM parameters are not
very quantitative. Theoretical uncertainties are greatly reduced in
the ratios of the branching ratios involving the decays $B \to (\rho, \omega)
\gamma$ and $B \to K^* \gamma$. Calling the ratios of the branching ratios
$R^\pm (\rho \gamma/K^* \gamma)$ and $R^0 (\rho \gamma/K^* \gamma) $,
for the decays of the $B^\pm$ and $B^0$ mesons, respectively, one has~\cite{Ali:2001ez}
\begin{eqnarray}
R^\pm (\rho \gamma/K^* \gamma) &=&  \left| V_{td} \over V_{ts} \right|^2
 \frac{(M_B^2 - M_\rho^2)^3}{(M_B^2 - M_{K^*}^2)^3}~\zeta^2
(1 + \Delta R^\pm(\epsilon_{\rm A}^\pm, \bar{\rho}, \bar{\eta})) \; , 
\nonumber\\[-1.5mm] 
\label{rapp}\\[-1.5mm] 
R^0 (\rho \gamma/K^* \gamma) &=& {1\over 2} \left| V_{td}
 \over V_{ts} \right|^2
 \frac{(M_B^2 - M_\rho^2)^3}{(M_B^2 - M_{K^*}^2)^3}~\zeta^2
(1 + \Delta R^0 (\epsilon_{\rm A}^0, \bar{\rho}, \bar{\eta})) \; ,
\nonumber
\end{eqnarray}
where
$\zeta=T_1^{\rho}(0)/T_1^{K^*}(0)$, with $T_1^{\rho}(0)$ and $T_1^{K^*}(0)$
being the transition form factors evaluated at $q^2=0$ in the decays $B \to \rho \gamma$ 
and $B \to K^* \gamma$, respectively. The functions
 $\Delta R^\pm (\epsilon_{\rm A}^\pm, \bar{\rho}, \bar{\eta})$
 and $\Delta R^0 (\epsilon_{\rm A}^0, \bar{\rho}, \bar{\eta})$, 
appearing on the r.h.s. of the above equations encode
both the $O(\alpha_s)$ contribution to the penguin
amplitudes and annihilation contributions, with the latter estimated so far only in
the lowest order. They  have a non-trivial dependence on the CKM 
parameters $\bar{\rho}$ and $\bar{\eta}$~\cite{Ali:2001ez,Bosch:2001gv}.
Theoretical uncertainty in the evaluation of the ratios $R^\pm (\rho \gamma /K^* 
\gamma)$ and $R^0 (\rho \gamma /K^* \gamma)$ is dominated by the error on
 the quantity $\zeta$, and to some extent also by the errors on the parameters
$\epsilon^\pm_A$ and $\epsilon^0_A$, characterizing the annihilation/penguin ratios.
 In the SU(3) limit $\zeta=1$; SU(3)-breaking corrections 
have been calculated in several approaches, including the QCD sum rules and Lattice QCD.
With the current values for the ratios $ R^\pm (\rho \gamma/K^* \gamma)=0.032
\pm 0.008$,  $R^0 (\rho \gamma/K^* \gamma)=0.039 \pm 0.007$ and
$R^0 (\omega \gamma/K^* \gamma)=0.021 \pm 0.007$, the current world average 
of $\vert V_{td}/V_{ts}\vert$ from the ratio of $B \to (\rho,\omega)\gamma$
and $B \to K^* \gamma$ is~\cite{Koneke:2007eu}:
\begin{equation}
\vert V_{td}/V_{ts}\vert =0.194 ^{+0.015}_{-0.014}({\rm exp}) \pm 0.014 ({\rm th})~, 
\end{equation}
where the Light-cone QCD sum rules~\cite{Ball:2006nr} have been used  
to estimate the hadronic input quantities. This determination is compatible with
the one from  the mass difference ratio $\Delta M_{B_s}/\Delta M_{B_d}$
given in eq.~(\ref{eq:vtdts-deltas}), but less precise.

\section{$B \to M_1 M_2$ Decays}
Exclusive non-leptonic decays are the hardest nuts to crack 
in the theory of $B$-decays.
Basically, there are four different theoretical approaches 
to calculate and/or parameterize the hadronic matrix elements 
in $B \to M_1 M_2$ decays:
\begin{enumerate}
\item SU(2)/SU(3) symmetries and phenomenological Ansaetze
~\cite{Gronau:1990ka,Buras:1994pb,Deshpande:1994pw,Gronau:2002gj}

\item Dynamical approaches based on perturbative QCD, 
such as the QCD Factorization~\cite{Beneke:1999br}
and the competing pQCD approach~\cite{Keum:2000ph}. 

\item Charming Penguins~\cite{Ciuchini:2001gv} 
using the renormalization group invariant topological approach 
of Buras and Silvestrini~\cite{Buras:1998ra}.
 
\item Soft Collinear Effective Theory 
(SCET)~\cite{Bauer:2000ew,Bauer:2000yr,Bauer:2001yt,Beneke:2002ph},
which we have already discussed
in the context of radiative and semileptonic decays.
\end{enumerate}

These approaches will be discussed on the example of the 
$B \to \pi \pi$ and $B \to K\pi$ decays for which now there 
exist enough data to extract the underlying dynamical parameters.
Prior to this, however, we discuss the measurements of the angle
$\beta$ (or $\phi_1$) from the experiments at the $B$-factories.

\subsection{Interplay of Mixing and Decays 
      of $B^0$- and $\bar B^0$-Mesons to CP Eigenstates} 
We start with the discussion of the transition $b \to c c \bar s$, which is
dominated by the tree topology. The time-dependent CP asymmetry in the decays
$B^0 \to f$ and $\bar {B}^0 \to f$, where $f$ is a CP eigenstate, such as
$J/\psi K_s$ and $J/\psi K_L$, is defined as:
\begin{equation} 
{\cal A}_f (t) = 
\frac{\Gamma[\bar B^0 (t) \to f] - \Gamma[B^0 (t) \to f]}
     {\Gamma[\bar B^0 (t) \to f] + \Gamma[B^0 (t) \to f]} . 
\label{eq:ACP-time-def}
\end{equation}
The time evolution of the two flavor eigenstates  $B^0$  and $\bar{B}^0$
is determined by  
 $(2 \times 2)$ Hermitian matrices $M$ and $\Gamma$.
The physical states (with definite masses and lifetimes) are the linear
combinations
of  $B^0$  and $\bar{B}^0$, with $|B_d(L,H)\rangle = p|B^0\rangle \pm q|\bar{B}^0\rangle$,
 dependent on two complex parameters~$p$ and~$q$. Defining the decay 
amplitudes $A(f) \equiv \langle f | H | B^0 \rangle$  and 
$\bar A (f) \equiv \langle f | H | \bar B^0 \rangle$ of the $B^0$- and 
$\bar B^0$-mesons into the final state~$f$, the time-dependent CP asymmetry is
determined by the quantity $\lambda_f$, involving the interplay of mixing and
decay amplitudes:
\begin{equation} 
\lambda_f = \frac{q}{p} \, \rho(f),  
\qquad 
\rho(f) = \frac{\bar A (f)}{A (f)}. 
\label{eq:lambda-f}
\end{equation}
For the  $B_d^0$ - $\bar{B}_d^0$ mixing, the ratio $q/p$ involves the phase
$\beta$
(or $\phi_1$), which is one of the angles of the unitarity triangle:
\begin{equation} 
\frac{q}{p} = \frac{V^*_{tb} V_{td}}{V_{tb} V^*_{td}}
= {\rm e}^{- 2 i \phi_{\rm mixing}} = {\rm e}^{- 2 i \beta}.  
\end{equation}
The time dependent CP asymmetry~(\ref{eq:ACP-time-def}) is then expressed
as
\begin{equation} 
{\cal A}_f (t) = C_f \cos (\Delta M_{B_d} t) + S_f \sin (\Delta M_{B_d} t), 
\end{equation}
where $\Delta M_{B_d} = (0.507 \pm 0.005)$~ps$^{-1}$ is the mass 
difference between the heavy and light $B_d^0$-meson mass eigenstates 
and the difference in the decay widths $\Delta \Gamma_{B_d}$ has been neglected. 
The quantities~$C_f$ and~$S_f$, called the direct and mixing-induced 
CP asymmetries, respectively, are defined in terms of the complex
variable $\lambda_f$ as follows: 
\begin{equation} 
C_f = \frac{1 - |\lambda_f|^2}{1 + |\lambda_f|^2} , 
\qquad 
S_f = \frac{2 {\rm Im} \lambda_f}{1 + |\lambda_f|^2}. 
\label{eq:Cf-Sf-def}
\end{equation}
If the decays $B^0 \to f$ and $\bar{B}^0 \to f$ are dominated by a {\it single}
amplitude, the ratio $\rho(f) = \eta_f \, 
{\rm e}^{- 2 i \phi_{\rm decay}}$, where $\eta_f = \pm 1$ is 
the CP parity of the state~$f$, is a pure phase factor and the
asymmetries~(\ref{eq:Cf-Sf-def}) reduce to the expressions: 
\begin{equation} 
C_f = 0, 
\qquad 
S_f = - \eta_f \sin 2 (\phi_{\rm mixing} + \phi_{\rm decay}) . 
\label{eq:Cf-Sf-SingleAmpl}
\end{equation}
The decays 
$B^0/\bar{B}^0 \to J/\psi K_s, J/\psi K_L$, and a number of related
final states with  $f$ being $\psi(2S) K_s$, $\eta_c K_s$, $\chi_{c1} K_s$,
 and $J/\psi K^{*0} 
(K^{*0} \to K_s \pi^0)$ belong to the category of
 {\it gold plated} decays~\cite{Carter:1980tk}.
In all these modes, the direct CP asymmetry $C_f$, to a very high accuracy,
vanishes, and the quantity $S_f$, 
the mixing-induced CP asymmetry, measures $\sin (2 \beta)$. 
Averaging over all the  decay channels, the results of the BABAR and
BELLE measurements are as follows~\cite{Brown-LP07}: 
\begin{eqnarray} 
&& \hspace*{-10mm}
C =  0.049 \pm 0.022 \pm 0.017, \quad  
S =  0.714 \pm 0.032 \pm 0.018, \quad 
[{\rm BABAR}] 
\label{eq:BABAR-res} \\ 
&& \hspace*{-10mm} 
C = -0.019 \pm 0.025, \hspace*{13mm} 
S =  0.651 \pm 0.034. \hspace*{17mm} 
[{\rm BELLE}]  
\label{eq:BELLE-res}
\end{eqnarray}
In the BABAR result,  the first error is statistical and 
the second is systematic, while in the BELLE data both the errors
have been combined.  The current world average for $S_f=\sin (2 \beta)$
for the quark transition $b \to c \bar c s$ is~\cite{hfag07}:
\begin{equation}
\sin (2 \beta) = 0.681 \pm 0.025,  
\label{eq:HFAG-res}
\end{equation}
where the data from LEP and Tevatron have also been included. 
Restricting $\beta$ in the range $0 \le \beta \le \pi/2$, two 
possible values can be extracted $\beta = (21.5 \pm 1.0)^\circ$ 
and  $\beta = (68.5 \pm 1.0)^\circ$. The two-fold ambiguity has now been
resolved by several $\cos (2 \beta)$ measurements, involving the Dalitz analysis
of the decay modes $B^0 \to D^0_{\rm 3-body} h^0$, $ B^0 \to K_s \pi^+\pi^-$,
$B^0 \to K_s K^+ K^-$, and the older results on $B^0 \to J/\psi K^{*0}$, leading
to the determination  $\beta = (21.5 \pm 1.0)^\circ$.

The {\it direct} measurement of $\sin (2 \beta)$ in eq.(\ref{eq:HFAG-res}) is to
  be compared with the {\it indirect} estimate of the same, obtained from the
fits of the CKM unitarity triangle (UT). For this, the UTfit
 collaboration~\cite{Bona:2007vi}
quotes  $\sin (2 \beta)= 0.739 \pm 0.044$, obtained from the sides of the 
UT alone, and  $\sin (2 \beta)= 0.736 \pm 0.042$, by including also the
CP-violating  quantity $\epsilon_K$ in the $K$-decays. The results from the
CKMfitter group~\cite{Charles:2004jd} are similar. Thus, SM passes this test
comfortably. 

Another key test of the SM in the flavor sector is to compare the CP-violating
quantities $S_f$ and $C_f$ involving the penguin-topology dominated quark
transitions $b \to s \bar{s} s$ and $b \to s d \bar{d}$ with the ones from
the transition $b \to c \bar{c} s$, dominated by the tree topology and
discussed quantitatively above. The point here is that penguin amplitudes may
receive contributions from New Physics. For example, new phases, present
generically
in supersymmetric theories, may reveal themselves, leading to $S_{f=c\bar{c}s}
\neq S_{f=s \bar{s} s; sd\bar{d}}$. Examples of the final states induced by
the transition $b \to s \bar{s} s$ are $(\phi, \eta, \eta^\prime, K\bar{K}) K_s$,
and the ones induced by the transition $b \to s d \bar{d}$ are
$K_s^0 (\pi^0, \rho^0, \omega)$ ($B^0 \to \eta^\prime K_s$ receives
contributions
from both the transitions). The current measurements of $S_f=- \eta_f \sin
(2\beta^{\rm eff})$ from the penguin-dominated decays are~\cite{hfag07}:
 $S_{\phi K^0}=0.39 \pm 0.17$,
$S_{\eta^\prime K^0}=0.61 \pm 0.07$, $S_{K_s K_s K_s}=0.58 \pm 0.20$,
$S_{\pi^0 K_s}=0.38 \pm 0.19$, $S_{\rho^0 K_s}=0.61 ^{+0.25}_{-0.27}$,
$S_{\omega K_s}=0.48 \pm 0.24$, $S_{f_0 K^0}=0.84 \pm 0.07$,
$S_{K^+K^-K^0}=0.73\pm 0.10$, and  $S_{\pi^0\pi^0 K_s}=-0.52
\pm 0.41$. These measurements are not as precise as the ones from the
$b \to c \bar{c}s$ decays due to the much smaller branching ratios (typically
$10^{-5}$) compared to the decay $B^0 \to J/\psi K_s$.
Also, they involve more than one decay topologies. However, within (large)
errors,  the  values of $\sin(2\beta^{\rm eff})$ from the penguin-dominated transitions
are consistent with the value of $\sin (2 \beta)$ from the
tree-dominated transition given above in eq.(\ref{eq:HFAG-res}), with the
possible exception
of $S_{\phi K^0}$, which deviates by about 2$\sigma$, and the poorly
measured odd-man out $S_{\pi^0\pi^0 K_s}$. It seems that the fog on $S_f$ in the
penguin-dominated decays from the initial epoch of the $B$-factory experiments
has largely evaporated, and the emerging contours of CP asymmetries in these
decays are very much the same as predicted by the SM. 

\subsection{$B \to \pi \pi$: SM vs.~Experiments}
The determination of the phase $\alpha$ is based on the branching ratios and
CP asymmetries in the quark transition $b \to u \bar{u} d$. They metamorphise
in the decays $B \to \pi \pi$, $B \to \rho \pi$ and $B \to \rho \rho$, apart
from other final states. We concentrate here on the decay $B \to \pi \pi$,
which has received a lot of theoretical attention.
 
There are three dominant topologies in the $B \to \pi \pi$ 
decays termed as Tree (T), Penguin (P) and Color-suppressed (C). 
In addition, there are several other subdominant topologies 
which will be neglected in the discussion below. Parametrization 
of the T, P, and C amplitudes is convention-dependent. In the 
Gronau-Rosner c-convention~\cite{Gronau:2002gj}, these amplitudes 
can be represented as
\begin{eqnarray}
\sqrt 2 \, A^{+0} & = &  
- |T| \, {\rm e}^{i \delta_T} \, {\rm e}^{i \gamma} 
\left [ 1 + |C/T| \, {\rm e}^{i \Delta} \right ]~, 
\nonumber \\ 
A^{+-} & = &  
- |T| \, {\rm e}^{i \delta_T} \left [ {\rm e}^{i \gamma} 
+ |P/T| \, {\rm e}^{i \delta} \right ]~, 
\label{eq:B-pipi-Ampl} \\  
\sqrt 2 \, A^{00} & = &  
- |T| \, {\rm e}^{i \delta_T} \left [ |C/T| \, 
{\rm e}^{i \Delta} \, {\rm e}^{i \gamma} 
- |P/T| \, {\rm e}^{i \delta} \right ]. 
\nonumber 
\end{eqnarray} 
The charged-conjugate amplitudes  $\bar A^{ij}$ 
differ by the replacement  $\gamma \to - \gamma$. 
The amplitudes~(\ref{eq:B-pipi-Ampl}) and the 
charged-conjugate ones obey the isospin relations: 
\begin{equation} 
A^{+0} = \frac{1}{\sqrt 2} \, A^{+-} + A^{00}, 
\qquad 
\bar A^{-0} = \frac{1}{\sqrt 2} \, \bar A^{+-} + \bar A^{00} . 
\label{eq:isospin-rel}  
\end{equation}

There are 5 dynamical parameters $|T|$, 
$r \equiv |P/T|$, $\delta$, $|C/T|$, $\Delta$, with 
$\delta_T = 0$ assumed for the overall phase. Thus, 
the weak phase~$\gamma$ can be extracted together 
with other quantities if the  complete set of 
experimental data on $B \to \pi\pi$ decays is available. 

\begin{table}[!h]
\tbl{Branching ratios (in units of $10^{-6}$) and CP asymmetries
 in the $B \to \pi \pi$ decays}
{\begin{tabular}{ll} \hline \\[-1mm]
${\cal B} (B^+ \to \pi^+ \pi^0) = 5.59^{+0.41}_{-0.40}$ \hspace*{11mm} & 
$A_{\rm CP}(\pi^+\pi^0) = 0.06 \pm 0.05$ \\[3mm] 
${\cal B} (B^0 \to \pi^0 \pi^0) = 1.31 \pm 0.21$ & 
$A_{\rm CP}(\pi^0\pi^0) = 0.48^{+0.32}_{-0.31}$ \\ [3mm] 
${\cal B} (B^0 \to \pi^+\pi^-) = 5.16 \pm 0.22$ & 
$C_{\rm CP}(\pi^+\pi^-) = -0.38 \pm 0.07$ \\[1mm]  
& 
$S_{\rm CP}(\pi^+\pi^-) = -0.61 \pm 0.08$ \\[2mm] \hline 
\end{tabular}}
\label{tab:B-pipi-exp}
\end{table} 
%

%
%
The experimental branching ratios and the direct CP asymmetries 
$A_{\rm CP}(\pi^0\pi^0)$ and $C_{\rm CP}(\pi^+\pi^-)$, as well 
as the value of the coefficient $S_{\rm CP}(\pi^+\pi^-)$ in 
time-dependent CP asymmetry, presented in Table~\ref{tab:B-pipi-exp}, 
have been fitted to determine the various parameters
(the direct CP asymmetry $A_{\rm CP}(\pi^+\pi^0)$ is not relevant 
for this analysis but can be important in determining the size 
of electroweak contribution in the decays considered). 
An updated analysis by Parkhomenko 
based on the paper~\cite{Ali:2004hb} yields the following values
for the hadronic parameters:
\begin{eqnarray}
&&
\left\vert P/T \right\vert = 0.473^{+0.060}_{-0.055}, 
\qquad  
\delta = (-40.2^{+6.8}_{-4.7})^\circ~,
\label{eq:B-pipi-HP} \\ 
&& 
\left\vert C/T \right\vert = 0.966^{+0.058}_{-0.061}, 
\qquad 
\Delta = (-56.3^{+8.4}_{-7.9})^\circ , 
\nonumber
\end{eqnarray} 
and for the CKM unitarity triangle angle~$\gamma$ 
(or equivalently~$\alpha$) 
\begin{equation} 
\gamma = \left ( 65.9^{+3.0}_{-3.2} \right )^\circ , 
\qquad 
\alpha = \pi - \beta -\gamma = \left ( 92.6^{+3.4}_{-3.2} \right )^\circ. 
\label{eq:alpha-gamma-fit}
\end{equation}
Similar fits based on their data have been performed by 
the BABAR and BELLE collaborations resulting in  slightly 
larger values: $\alpha = (96^{+10}_{-6})^\circ$ (BABAR) and 
$\alpha = (97 \pm 11)^\circ$ (BELLE). The overall 
fits performed by the CKM-Fitter and UT-Fit groups prefer slightly 
smaller values, yielding:  $\alpha = (90.7^{+4.5}_{-2.9})^\circ$ ~\cite{Charles:2004jd} 
and $\alpha = (88.7 \pm 6.2)^\circ$~\cite{Bona:2007vi}, respectively.
All the above estimates are in good agreement 
with each other within the quoted errors, stating that the data on
$B \to \pi \pi$ (as well as the other decay modes $B \to \rho \pi$ and $B \to
\rho \rho$) are in agreement  
 with the indirect estimate of the phase $\alpha$ from the unitarity triangle. 
The strong phases~$\delta$ and~$\Delta$ in Eq.(\ref{eq:B-pipi-HP}) 
come out rather large. In particular, they are much larger than the
predictions of the QCD-F 
approach~\cite{Beneke:1999br}, with pQCD~\cite{Keum:2000ph} 
in better agreement with data, but neither of these approaches provides 
a good fit of the entire $B \to \pi\pi$ data.

Data on $B \to \pi\pi$ decays are in agreement with the phenomenological 
approach of the so-called charming penguins~\cite{Ciuchini:2004tr}, 
and with the SCET-based analyses by
 Bauer et al.~\cite{Bauer:2004tj,Bauer:2005kd}  
which also attributes a dominant role to the charming penguin amplitude. 
However, a proof of the factorization of the charming penguin amplitude 
in the SCET approach remains to be provided. In addition, SCET makes 
a number of predictions in the $B \to \pi\pi$ sector, such as the 
branching ratio ${\cal B} (B^0 \to \pi^0\pi^0)$:~\cite{Bauer:2004tj}
\begin{equation}
 {\cal B} (B^0 \to \pi^0\pi^0) 
\bigg |_{\gamma = 64^\circ} \!\!\! = 
(1.3 \pm 0.6) \times 10^{-6}~. 
\end{equation}
In contrast,  predictions of the QCD-F and pQCD approaches are rather 
similar: $ {\cal B} (B^0 \to \pi^0\pi^0) \sim  0.3 \times 10^{-6}$, 
in substantial disagreement with the data.

\subsection{Present bounds on the phase $\gamma$ from $B$ decays}
The classic method for determining the phase $\gamma$ (or 
$\phi_3$)~\cite{Gronau:1991dp,Gronau:1990ra,Atwood:1996ci,Atwood:2000ck}
involves the interference of the tree amplitudes 
$b \to u W^- \to u \bar{c}s$ leading to $B^- \to D^0 K^-$ and $b 
\to c W^- \to c \bar{u} s$ leading to $B^- \to \overline{D^0} K^-$. 
These amplitudes can interfere
if $D^0$ and $\overline{D^0}$ decay into a common hadronic final state. 
Noting that the CP$=\pm 1$
eigenstates $D^0_{\pm}$ are linear combinations of the $D^0$ and 
$\overline{D^0}$ states: $D^0_{\pm}= (D^0 \pm \overline{D^0})/\sqrt{2}$,
both branches lead to the same final states $B^- \to D^0_{\pm} K^-$. So, the
condition of CP interferometry is fulfilled. The
decays $B^- \to D^0_{\pm} K^-$ are described by the amplitudes:
\begin{equation}
A(B^- \to D_{\pm}^0 K^-) = \frac{1}{\sqrt{2}}\,\left [ A(B^- \to D^0 K^-) \pm 
A(B^- \to \overline{D^0} K^-) \right ]\,.
\end{equation}
Since, the weak phase of the $b \to u$ transition is 
$\gamma$ but the $b \to c$ 
transition has no phase, a measurement of the CP asymmetry through the 
interference of these two amplitudes yields $\gamma$.
The four equations that will be used to extract $\gamma$ are:
\begin{eqnarray}
 R_{\pm}&\equiv& \frac{ {\cal B}(B^- \to D_{\pm}^0 K^-) + {\cal B}(B^+ \to 
D_{\pm}^0 K^+)}{{\cal B}(B^- \to D^0 K^-) + {\cal B}(B^+ \to D^0 K^+)}
= 1 + r_{\rm DK}^2 \pm 2 r_{\rm DK} \cos \delta_{\rm DK} \cos \gamma\,, 
\nonumber\\[-1mm]
\label{Lyuba} \\[-1mm]
A_{\pm} &\equiv& \frac{ {\cal B}(B^- \to D_{\pm}^0 K^-) - {\cal B}(B^+ \to 
D_{\pm}^0 K^+)}{ {\cal B}(B^- \to D_{\pm}^0 K^-) + {\cal B}(B^+ \to D_{\pm}^0 K^+)}
= \frac{\pm 2 r_{\rm DK} \sin \delta_{\rm DK} \sin \gamma}{1 +r_{\rm DK}^2 \pm 
2 r_{\rm DK} \cos \delta_{\rm DK} \cos \gamma}\,.
\nonumber
\label{dkasym}
\end{eqnarray}
Here, $r_{\rm DK}$ is the ratio of
the two tree amplitudes~\cite{Gronau:2002mu} 
$r_{\rm DK}\equiv \vert T_1/T_2 \vert  \sim (0.1 - 0.2)$, 
with $T_1$ and $T_2$ being the CKM suppressed $(b \to u)$ and CKM 
allowed $(b \to c)$ amplitudes, respectively, and 
$\delta_{\rm DK}$ is the relative strong phase between them.
The construction 
of the final states involves flavor and CP-tagging of the various $D^0$ 
states, which can be done, for example, through the decays 
$D^0_{+} \to \pi^+ \pi^-$,
$D^0_{-} \to K_S \pi^0$, and $D^0 \to K^- \pi^+$. With three unknowns
 ($r_{\rm DK}, \delta_{\rm DK}, \gamma)$, but four
quantities which will be measured, $R_{\pm}$ and $A_{\pm}$, one has, in 
principle, an over constrained system.

Experimentally, the quantities $R_{\pm}$ are measured through the ratios:
\begin{eqnarray}
R(K/\pi) &\equiv & \frac{{\cal B}(B^- \to D^0 K^-)}
{{\cal B}(B^- \to D^0 \pi^-)}\,, 
\qquad
R(K/\pi)_{\pm} \equiv \frac{{\cal B}(B^\pm \to D^0_{\pm} K^\pm)}
{{\cal B}(B^\pm \to D^0_{\pm} \pi^\pm)}\,.
\label{rkpipm}
\end{eqnarray}

With all three quantities $R(K/\pi)$ and $R(K/\pi)_{\pm}$ measured, 
one can determine  
$R_{\pm} = R(K/\pi)_{\pm}/R(K/\pi)$.
More useful decay modes to construct 
the $B \to DK$ triangle can be added  
 to reduce the statistical errors. Along these lines, Atwood and
Soni~\cite{Atwood:2003jb} have advocated
to also include the decays of the vector states in the analysis, 
such as $B^- \to K^{*-} D^0$, $B^- \to K^- D^{*0}$, and 
$B^- \to K^{*-} D^{*0}$, making use of the $D^{*0} \to D^0
\gamma$ and $D^{*0} \to D^0 \pi^0$ modes.

 Present measurements in the $B \to 
DK$ and $B \to D \pi$ decays by the BABAR and BELLE collaborations yielding
 $R_{\pm}$ and $A_{\pm}$ for the $D_{\rm  CP} K^-$ mode are summarized by
 HFAG~\cite{hfag07}: 
\begin{eqnarray}
R_{+} &=& 1.09 \pm 0.09\,, \qquad 
A_{+}= 0.26 \pm 0.08 ~~{\rm [BELLE, BABAR]}\,, 
\nonumber\\
R_{-} &=& 0.90 \pm 0.10\,, \qquad 
A_{-}= -0.16 \pm 0.09 ~~{\rm [BELLE]}\,. 
\label{gammadk} 
\nonumber
\end{eqnarray}
The corresponding quantities for the  $D_{\rm  CP}^* K^-$ and
 $D_{\rm  CP} K^{*-}$ are also given by  HFAG~\cite{hfag07}.

A modification of the Gronau-London-Wyler (GLW) method described above  has been suggested by Atwood, Dunietz and
Soni (ADS), where $B^- \to D^0K^-$ with $D^0 \to K^+\pi^-$ (or similar) and the
charge conjugate decays are implied. BABAR and BELLE use the following
definitions
for the quantities called $A_{\rm ADS}$ and $R_{\rm ADS}$, (the decay modes
$B^- \to D^0K^-$ followed by $D^0 \to K^+\pi^-$ are used to exemplify the method)
\begin{eqnarray}
 R_{\rm ADS}&\equiv& \frac{ {\cal B}(B^- \to [K^+ \pi^-]_D K^-) + {\cal B}(B^+ \to 
[K^-\pi^+]_D K^+)}{{\cal B}(B^- \to [K^-\pi^+]_D K^-) + {\cal B}(B^+ \to [K^+\pi^-]_D K^+)}
\,, \nonumber\\[-1mm]
\label{Lyuba-2} \\[-1mm]
A_{\rm ADS} &\equiv& \frac{ {\cal B}(B^- \to [K^+\pi^-]_D K^-) - {\cal B}(B^+ \to 
[K^-\pi^+]_D K^+)}{ {\cal B}(B^- \to [K^+\pi^-]_D K^-) + {\cal B}(B^+ \to
[K^-\pi^+]_D K^+)}\,.
\nonumber
\label{dkasym-2}
\end{eqnarray}
The current measurements of these observables are summarized by HFAG~\cite{hfag07}.
 In the analysis of data, usually the GLW and ADS methods are
combined and a $\chi^2$-fit is done to determine the profile of the phase $\gamma$.

A variant of the $B \to DK$ method of measuring $\gamma$ is to use the decays
$B^\pm \to DK^\pm$ followed by  multi-body decays of the $D$-meson, such as
$D^0 \to K_S \pi ^- \pi^+$, $D^0 \to K_S K^- K^+$ and $D^0 \to K_S \pi^- \pi^+ 
\pi^0$, in which a binned Dalitz plot analysis  of the decays 
$D^0/\overline{D^0} \to K_S \pi^- \pi^+$ 
was proposed~\cite{Atwood:2000ck,Giri:2003ty}.
 Assuming no CP asymmetry in $D^0$ decays, the amplitude of the
$B^+ \to D^0 K^+ \to (K_S \pi^+ \pi^-) K^+$ can be written as
\begin{equation}
M_{+}= f(m_+^2,m_-^2) + r_{DK} {\rm e}^{i( \gamma + \delta_{DK})} f(m_-^2, m_+^2)\,,
\label{dalitz1}
\end{equation}
where $m_+^2$ and $m_-^2$ are the squared invariant masses of the $K_S\pi^+$ and 
$K_S \pi^-$ combinations in the $D^0$ decay, and $f$ is the complex amplitude of
the decay $D^0 \to K_S \pi^+ \pi^-$. The quantities $r_{DK}$ and $\delta_{DK}$ are 
the relative magnitudes and  strong phases of the two amplitudes, already 
discussed earlier. The amplitude for the charge conjugate $B^-$ decay is
\begin{equation}
M_{-}= f(m_-^2,m_+^2) + r_{DK} {\rm e}^{i(- \gamma + \delta_{DK})} f(m_+^2, m_-^2)\,.
\label{dalitz2}
\end{equation}
Once the functional form of $f$ is fixed by a choice of a model 
for $D^0 \to K_S \pi^+ \pi^-$ decay, the Dalitz distribution for 
$B^+$ and $B^-$ decays can be fitted simultaneously by the expressions 
for $M_+$ and $M_-$, with $r_{DK}$, $\delta_{DK}$
and $\gamma$ (or $\phi_3$) as free parameters. The model-dependence could 
be removed by a binned Dalitz~distribution~\cite{Giri:2003ty}.
This is usually called the GGSZ method, and has been used to determine
$\gamma$. The combined fit of both of these methods by
CKMfitter~\cite{Charles:2004jd} yields $\gamma =(76.8^{+30.4}_{-31.5})^\circ$,
to be compared with their overall fit from the CKM unitarity
$\gamma =(67.6^{+2.8}_{-4.5})^\circ$. The corresponding fit by the UTfit
 group~\cite{Bona:2007vi} yields $\gamma =(67 \pm 7)^\circ$. Thus, we see that
within the current experimental error of the {\it direct} measurements, which
is quite large, also the phase $\gamma$
is compatible with its {\it indirect} estimates in the SM. The experimental
precision will greatly improve at the LHC, in particular, by using 2-body
$B_s$-decays. 

\subsection{$B \to K \pi$: SM vs.~Experiments}
We now discuss the decays $B \to K \pi$. 
First, we note that the direct CP-asymmetry in the $B \to K\pi$ 
decays has now been measured by the BABAR, BELLE and CDF collaborations:
\begin{equation}
A_{\rm CP}(\pi^+K^-) = 
\left \{ 
\begin{array}{lc}
(-10.7 \pm 1.8^{+0.7}_{-0.4})\% & [{\rm BABAR}], \\ 
(-9.3 \pm 1.8 \pm 0.8)\% & [{\rm BELLE}], \\
(-8.6 \pm 2.3 \pm 0.9)\% & [{\rm CDF}],  
\end{array} 
\right. 
\label{eq:ACP-piK-exp} 
\end{equation} 
to be compared with the predictions of the two
factorization-based approaches: 
$A_{\rm CP}(\pi^+K^-) = (-12.9 \div -21.9)\% 
[{\rm pQCD}]$~\cite{Keum:2000ph}
and $A_{\rm CP}(\pi^+K^-)= (-5.4 \div +13.6)\% 
[{\rm QCD-F}]$~\cite{Beneke:1999br}, with the latter 
falling short of a satisfactory description of data.

The charged and neutral $B \to \pi K$ decays have received 
a lot of theoretical attention. In particular, many ratios 
involving these decays have been proposed to test the 
SM~\cite{Fleischer:1997um,Neubert:1998jq,Lipkin:1998ie,Buras:2000gc}
and extract useful bounds on the angle~$\gamma$, starting 
from the Fleischer-Mannel bound~\cite{Fleischer:1997um}:  
%
\begin{equation}
\sin^2\gamma \leq R \equiv 
\frac {\tau_{B^+}}{\tau_{B_d^0}} 
\frac{{\cal B}(B_d^0 \to \pi^-K^+) + {\cal B}(\bar B_d^0 \to \pi^+K^-)}
     {{\cal B}(B^+ \to \pi^+K^0) + {\cal B}(B^- \to \pi^-\bar K^0)}~.    
\label{eq:FM-bound}
\end{equation}
The current experimental average $R = 0.899 \pm 0.049$ allows 
to put a bound: $\gamma < 92^\circ$ (at 95\% C.L.). This is 
in agreement with the determination of~$\gamma$ from the 
$B \to \pi \pi$ and $B \to D^{(*)} K^{(*)}$ decays given earlier and the indirect 
unitarity constraints. Thus, both~$R$ and~$A_{\rm CP}(\pi^+K^-)$ 
are in agreement with the SM. The same is the situation with 
the Lipkin sum rule~\cite{Lipkin:1998ie}:
\begin{equation}
 R_L \equiv  2 \, 
\frac{\Gamma (B^+ \to K^+\pi^0) + \Gamma(B^0 \to K^0\pi^0)}
     {\Gamma (B^+ \to K^0\pi^+) + \Gamma(B^0 \to K^+\pi^-)} 
= 1 + {\cal O}(\frac{P_{\rm EW} +T}{P})^2~, 
\end{equation} 
implying significant electroweak penguin contribution in case~$R_L$ 
deviates significantly from unit. With the current experimental 
average $R_L = 1.071 \pm 0.049$, this is obviously not the case.
This leaves then the two other ratios~$R_c$ and~$R_n$ involving 
the $B \to \pi K$ decays of $B^\pm$ and $B^0$ mesons:
\begin{equation}
R_c \equiv 2 \, 
\frac{{\cal B}(B^\pm \to \pi^0 K^\pm)}{{\cal B}(B^\pm \to \pi^\pm K^0)}~, 
\qquad 
R_n \equiv \frac{1}{2} \, 
\frac{{\cal B}(B_d^0 \to \pi^\mp K^\pm)}{{\cal B}(B_d^0 \to \pi^0 K^0)}~.
\end{equation}
Their experimental values $R_c = 1.117 \pm 0.071$ and 
$R_n = 0.980 \pm 0.067$ are to be compared with the current 
SM-based estimates~\cite{Buras:2004th} $R_c = 1.14 \pm 0.05$ 
and $R_n = 1.11 ^{+0.04}_{-0.05}$. This implies 
$R_c({\rm SM}) - R_c ({\rm Exp}) = 0.02 \pm 0.09$ and
$R_n({\rm SM}) - R_n ({\rm Exp}) = 0.13 \pm 0.08$. 
Possible deviations from the SM, if confirmed, would imply 
new physics, advocated in this context, in particular, by 
Yoshikawa~\cite{Yoshikawa:2003hb}, Beneke and Neubert~\cite{Beneke:2003zv}  
and  Buras et al.~\cite{Buras:2004th}. However, as of now, one has to
conclude  that SM is in agreement with the measurements 
of both~$R_c$ and~$R_n$. 

Finally, a bound on ${\cal B}(B^0 \to K^0 \overline{K^0})$
based on $SU(3)$ and $B \to \pi \pi$ data, obtained recently 
by Fleischer and Recksiegel~\cite{Fleischer:2004vu}, yielding 
${\cal B}(B^0 \to K^0 \bar K^0) < 1.5 \times 10^{-6}$ 
is well satisfied by the current world average~\cite{hfag07} 
${\cal B}(B^0 \to K^0 \bar K^0) = (0.96^{+0.21}_{-0.19}) \times 10^{-6}$.

\section{$B_s^0$ Physics: Eldorado for the Tevatron and the LHC}
The main goal of $b$ physics at the hadron colliders Tevatron and the LHC is
to chart out the physics of the $B_s^0$ and $B_c^\pm$ mesons and of the
$b$-baryons. The current information on the spectroscopic and decay
characteristics of these hadrons is still very much in the offing, though clearly the two
Tevatron experiments have made some incisive inroads in these otherwise
uncharted territories.  Despite the overwhelming performance
of the $B$ factory experiments, there still remain a few landmark measurements
to be carried out involving $B_d^0$ and $B^\pm$ mesons. These include,
precise measurements of the CP asymmetries in the penguin-dominated
exclusive decays, quantitative determinations of the 
Wilson coefficients in the effective theory for weak decays $(C_7, C_8, C_9, C_{10})$, which will be
made possible by the precise measurements of the radiative and semileptonic
decays $B \to (X_s, K^*) \gamma$ and $B \to (X_s, K, K^*) \ell^+ \ell^-$.
It is challenging to measure the inclusive decays at the LHC, but
certainly exclusive decays will be well measured.

 In this section, a brief
list of some selected $b$ physics topics to be studied at the LHC is given and
discussed.

\begin{itemize}
\item $B_s^0$ - $\overline{B_s^0}$ Mixing.
\end{itemize}
 Apart from the precise
  measurement of $\Delta M_{B_s}=(17.77 \pm 0.10 \pm 0.07)$ (ps)$^{-1}$ by the
 CDF collaboration, there are two
  other quantities still to be measured in this complex: Lifetime difference
  $\Delta \Gamma_{B_s}$ and the phase $\phi_s$.
  These quantities have been calculated to a high precision in the
 SM~\cite{Beneke:1998sy}. A recent
 update of this work yields~\cite{Lenz:2006hd}
\begin{eqnarray}
\frac{\Delta \Gamma_{B_s}}{\Delta M_{B_s}}&=& (49.7 \pm 9.4) \times 10^{-4};
\hspace*{3mm} \Delta \Gamma_{B_s}=(0.096 \pm 0.039) ({\rm ps})^{-1}, \nonumber\\
 \phi_s
&=& (4.2 \pm 1.4) \times 10^{-3} = 0.24^\circ \pm 0.08^\circ~.
\end{eqnarray} 
The current measurements of these quantities from the D0
 collaboration are~\cite{Abazov:2007tx}
\begin{equation}
\Delta \Gamma_{B_s}=(0.12 ^{+0.08}_{-0.10} \pm 0.02) ({\rm ps})^{-1}~({\rm
  assuming}~\phi_s=0)~,
\end{equation}
and
\begin{equation}
\Delta \Gamma_{B_s}=(0.17 \pm 0.09 \pm 0.02) ({\rm ps})^{-1}~;
 \phi_s =-0.79 \pm 0.56 ^{+0.14}_{-0.01}~.
\end{equation}
The corresponding measurement (assuming $\phi_s=0$) for $\Delta \Gamma_{B_s}$
 from CDF is~\cite{Kuhr:2007dt}
 $\Delta \Gamma_{B_s}=(0.076 ^{+0.059}_{-0.063} \pm 0.006)$ (ps)$^{-1}$,
where the first error is statistical and the second systematic. 
At the LHCb~\cite{LHCb-07}, one anticipates a statistical sensitivity of $\sigma(\sin \phi_s)
\sim 0.031$ and  $\sigma(\Delta \Gamma_s/\Gamma_s) \sim 0.011$, assuming an
integrated luminosity of 2 (fb)$^{-1}$ and using the decay $B_s \to J/\psi
\phi$. This sensitivity will be improved by accumulating more data and
adding the CP modes $B_s \to J/\psi \phi$ and the pure CP modes
$B_s \to J/\psi \eta$ and $B_s \to J/\psi \eta^\prime$. The ATLAS and CMS
sensitivities on $\phi_s$
are expected to be somewhat worse by typically a factor 2. 
 Thus, experiments at the LHC will be able to test the SM
estimates for both the quantities $\Delta \Gamma_{B_s}$ and $\phi_s$. 

\begin{itemize}
\item Precise measurement of the phase $\gamma$.
\end{itemize}
At the LHC one can combine the methods of determining the phase $\gamma$
already in use at the $B$ factories involving the decays $B^+ \to D^{(*)}K^+$ and $B^0 \to
DK^{(*)}$, with the decay $B_s^0 \to D_s K$. In addition, one can use the
U-spin symmetry arguments advocated, in particular, by Fleischer~\cite{Fleischer:2007wr}, to
combine data from the
decays $B_d^0 \to \pi^+\pi^-$ and $B_s^0 \to K^+K^-$ to constrain $\gamma$.
An educated guess~\cite{schneider:07} is that a
sensitivity $\sigma(\gamma)\simeq 4^\circ$ with 2 (fb)$^{-1}$ of data is
reachable at the
LHCb, improving to $\sigma(\gamma)\simeq 2.4^\circ$ with  10 (fb)$^{-1}$. This
will result in  an order of magnitude improvement over the current precision on this
phase. Modest improvements are also anticipated for the other two phases
$\alpha$ and $\beta$ at the LHC.
\begin{itemize}
\item Leptonic decay $B^0_s \to \mu^+\mu^-$
\end{itemize}
New and improved upper limits have been presented  by the
 CDF~\cite{Mack:2007ra} and
 D0~\cite{do:5344}
 collaborations
for the decays $B^0_s \to \mu^+\mu^-$ and $B^0_d \to \mu^+\mu^-$. They are as
 follows (at 95\% C.L.)
\begin{eqnarray}
{\cal B}(B^0_s \to \mu^+\mu^-) &<& 9.3~[5.8] \times 10^{-8} ~~{\rm D0 [CDF]}~,\nonumber\\&&\hspace*{-3.0cm}{\cal B}(B^0_d \to \mu^+\mu^-) < 1.8 \times 10^{-8}~~{\rm [CDF]}.
\end{eqnarray}
The CDF and DO upper limits have been combined to yield~\cite{Nakao-LP07} 
${\cal B}(B_s^0 \to \mu^+\mu^-) < 4.5 \times 10^{-8}$, to be compared with the
SM predictions~\cite{Buchalla:1993bv}
 ${\cal B}(B^0_s \to \mu^+\mu^-) = 3.4 \times 10^{-9}$
and ${\cal B}(B^0_d \to \mu^+\mu^-)= 1.0 \times 10^{-10}$ with
$\pm 15\%$ theoretical uncertainty.
Hence, currently there is no sensitivity for the SM decay rate.
 However, as the leptonic branching ratios
probe the Higgs sector in beyond-the-SM scenarios, such as supersymmetry, and
they depend sensitively on 
$\tan \beta$, the Tevatron upper limit on ${\cal B}(B^0_{s} \to \mu^+\mu^-)$
 probes the large $\tan \beta$ 
parameter space, though the precise constraints are model
 dependent~\cite{Dermisek:2003vn,Ellis:2004tc}.
 At the LHC, the two main collider experiments ATLAS and CMS will reach
 the SM sensitivity, certainly with the higher LHC luminosity, $L_{\rm
   LHC}=10^{34}~cm^{-2}$ s$^{-1}$, as the decay $B_s^0 \to \mu^+\mu^-$
 remains
triggerable with the high luminosity.

\begin{itemize}
\item Charmless non-leptonic $B_s \to h_1 h_2$ decays.
\end{itemize} 
The experimental program to study non-leptonic decays $B_s^0 \to
h_1h_2$ has  started (here $h_{1,2}$ stand for charmless light vector or
pseudoscalar mesons) with first measurements for the
branching ratios $B_s^0 \to K^+ \pi^- $ and
 $B_s^0 \to K^+ K^- $
made available recently by the CDF collaboration~\cite{Abulencia:2006mq,Morello:2006pv}.
 Remarkably, the first direct CP asymmetry involving the decay
$B_s^0\to K^+ \pi^-$  and its CP conjugate mode  reported by
CDF is found to be large, with $A_{\rm
CP}(\overline{B_s^0}\to K^+ \pi^- ) =(39 \pm 15 \pm 8)\%$.
This large CP asymmetry was predicted by Lipkin~\cite{Lipkin:2005pb}
based on SU(3) symmetry arguments. This already tests various dynamical
models, such as  QCDF~\cite{Beneke:2006hg}, SCET~\cite{SCETBs}
and  pQCD~\cite{Ali:2007ff}. 
 With the ongoing $b$-physics program at
the Tevatron, but, in particular, with the onset of the LHC,
we expect a wealth of data involving the decays
 of the hitherto less studied $B_s^0$ meson.
The charmless $B_s^0 \to h_1 h_2$ decays are also important
for the CP asymmetry studies and the determination of the inner
angles of the unitarity triangle. As already stated,
a number of charmless decays $B_s^0 \to h_1 h_2$ can be related to
the $B_d^0 \to h_1 h_2$  decays using SU(3) (or U-spin) symmetry,
and hence data on these decays can be combined to test the
underlying standard model and search for physics beyond the SM under
less (dynamical) model-dependent conditions.
 Anticipating the
experimental developments,
many studies have been devoted to the interesting charmless
 $B_s^0 \to h_1 h_2$ decays, waiting to be tested at the LHC.

%
%

\section{Summary and Outlook}
Summarizing, dedicated experiments carried out over several decades combined
with  progress in  theoretical techniques embedded in QCD have enabled
a precise determination of the CKM matrix
elements. The knowledge of the third row and the third column of $V_{\rm CKM}$
has come from  $b$-physics, which we discussed
at length in this review. Of these, precise determination of 
 $V_{cb}$ and $V_{ub}$ required good control over the perturbative and
 non-perturbative aspects of QCD. The current precision on the
direct determination of $V_{tb}$ from the decay $t \to bW$ is limited by 
statistics and this will vastly improve from the top quark studies at the LHC
and later at the ILC.  The determination of $V_{ts}$ and $V_{td}$ require not
only precise knowledge of QCD in $b$ decays but implicitly also the
assumption of the CKM unitarity, as they are determined from the
loop-induced $b \to s$ and $b \to d$ transitions. Their current best
measurement is through the  mass differences  $\Delta M_{B_s}$ and $\Delta
M_{B_d}$, and the precision on these matrix elements (typically 10\%) is
completely dominated by theory.
 A complementary determination of  $V_{td}$ and $V_{ts}$ is also at hand
 from the radiative penguin transitions $b \to (d,s) \gamma$ 
and the exclusive decays $B \to (K^*,\rho,\omega)\gamma$,  but the current precision
is limited by both experimental statistics and non-perturbative aspects of
QCD, such as the transition form factors in exclusive decays. 
This surely will improve over the next several years. 

Experiments have also firmly established the phenomenon of CP violation in the $K$
and $B$ meson sectors. The various CP asymmetries in these decays are found
 compatible with
each other and, with some help from QCD, have a consistent interpretation
in terms of the single complex phase of the CKM matrix.  Again, in principle,
there is ample room also for beyond-the-SM weak phases, which would lead to very
different patterns of CP asymmetries in the tree-dominated versus loop-dominated
transitions. This has not been borne out by experiments at the $B$ factories.
 While the current data is not equivocal on all the decay
channels, and the dynamical aspects of not all the measured $B$-meson decays are 
quantitatively understood, the experimental case for  the
extra weak-phases is rather weak. Ongoing experiments at the $B$ factories
are expected to significantly reduce the errors on the quantities $S_f$ and $C_f$ in
penguin-dominated decays to settle the issue of new weak phases in $B$ decays.

 From the foregoing one has
to tentatively conclude that the CKM paradigm is now firmly
 established as the dominant mechanism of flavor
transitions in the quark sector. Whether future experiments, such as at the
LHC and (Super) $B$  factories, will force us to modify this paradigm remains to be seen.
We expect on theoretical  grounds that there is New
 Physics, probably just around the corner, to solve the outstanding issue
 of the gauge hierarchy. The resolution of this problem together with the
 unification of the gauge couplings and the search of viable candidate(s) for dark
 matter requires a TeV scale New Physics. Assuming that
 supersymmetry is the most viable candidate for the impending 
New Physics to be discovered by experiments at the LHC, the central issue
in the LHC era would be to pin down the underlying flavor aspects of this theory.
However, if the current experimental trend is any indicator, then very likely
the New Physics will be of the  
minimal flavor violating type, or something akin to it. Verifying
this and, more importantly, quantifying the subdominant flavor structures in
supersymmetry (or its alternative) is the next task to which theory and experiment
have to gear themselves up.

{\bf Acknowledgments}

I would like to congratulate Professor Riazuddin and the faculty of the
National Centre for Physics in Islamabad on the
auspicious occasion of the inauguration of the new campus.
I have thoroughly enjoyed this symposium marking the inaugural and thank 
Professor Riazuddin and his staff for their kind hospitality.
Thanks are also due to Alexander Parkhomenko for critically
reading this manuscript and his helpful remarks.
  
\vspace*{-0.5cm}

\end{document}